\documentclass{ws-procs10x7}

\begin{document}

\title{Rare decays and CP violation beyond the Standard Model}

\author{Luca Silvestrini}

\address{INFN, Sez. di Roma, Dip. di Fisica, Univ. di Roma ``La
  Sapienza'',\\ P.le A. Moro, I-00185 Rome, Italy. E-mail:
  Luca.Silvestrini@roma1.infn.it} 

\twocolumn[\maketitle\abstract{ We review the status of rare decays
  and CP violation in extensions of the Standard Model.  We analyze
  the determination of the unitarity triangle and the
  model-independent constraints on new physics that can be derived
  from this analysis. We find stringent bounds on new contributions to
  $K - \bar K$ and $B_d - \bar B_d$ mixing, pointing either to models
  of minimal flavour violation or to models with new sources of
  flavour and CP violation in $b \to s$ transitions. We discuss the
  status of the universal unitarity triangle in minimal flavour
  violation, and study rare decays in this class of models. We then
  turn to supersymmetric models with nontrivial mixing between second
  and third generation squarks, discuss the present constraints on
  this mixing and analyze the possible effects on CP violation in $b
  \to s$ nonleptonic decays and on $B_s - \bar B_s$ mixing. We
  conclude presenting an outlook on Lepton-Photon 2009.}]

\section{Introduction}
\label{sec:intro}

The Standard Model (SM) of electroweak and strong interactions works
beautifully up to the highest energies presently explored at
colliders. However, there are several indications that it must be
embedded as an effective theory into a more complete model that
should, among other things, contain gravity, allow for gauge coupling
unification and provide a dark matter candidate and an efficient
mechanism for baryogenesis. This effective theory can be described by
the Lagrangian
\begin{displaymath}
  \mathcal{L}(M_W)=\Lambda^2 H^\dagger H + \mathcal{L}_{\mathrm{SM}} +
  \frac{1}{\Lambda} \mathcal{L}^5 +
  \frac{1}{\Lambda^2} \mathcal{L}^6 + \ldots\,,
\end{displaymath}
where the logarithmic dependence on the cutoff $\Lambda$ has been
neglected. Barring the possibility of a conspiracy between physics at
scales below and above $\Lambda$ to give an electroweak symmetry
breaking scale $M_w \ll \Lambda$, we assume that the cutoff lies close
to $M_w$. Then the power suppression of higher dimensional operators
is not too severe for $\mathcal{L}^{5,6}$ to produce sizable effects
in low-energy processes, provided that they do not compete with
tree-level SM contributions. Therefore, we should look for new physics
effects in quantities that in the SM are zero at the tree level and
are finite and calculable at the quantum level. Within the SM, such
quantities fall in two categories: i) electroweak precision
observables (protected by the electroweak symmetry) and ii) Flavour
Changing Neutral Currents (FCNC) (protected by the GIM mechanism). The
first category has been discussed by S. Dawson at this conference,
while the second will be analyzed here.

In the SM, all FCNC and CP violating processes are computable in terms
of quark masses and of the elements of the Cabibbo-Kobayashi-Maskawa (CKM)
matrix. This implies very strong correlations among observables in the
flavour sector. NP contributions, or equivalently the operators in
$\mathcal{L}^{5,6}$, violate in general these correlations, so that NP
can be strongly constrained by combining all the available
experimental information on flavour and CP violation.

\section{The UT analysis beyond the SM}
\label{sec:UTA}

\begin{figure}[t]
\epsfxsize 0.49\textwidth
\figurebox{}{}{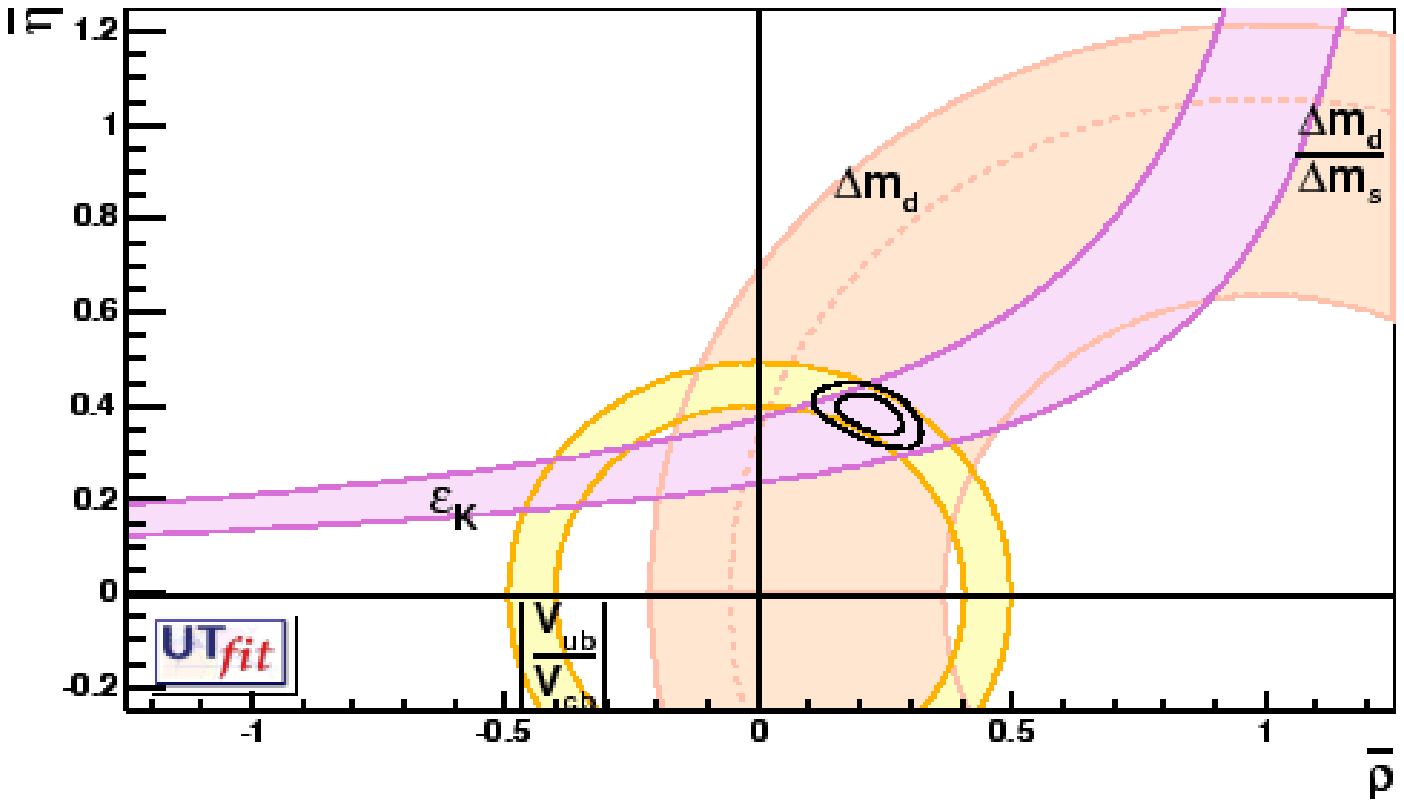}
\epsfxsize 0.49\textwidth
\figurebox{}{}{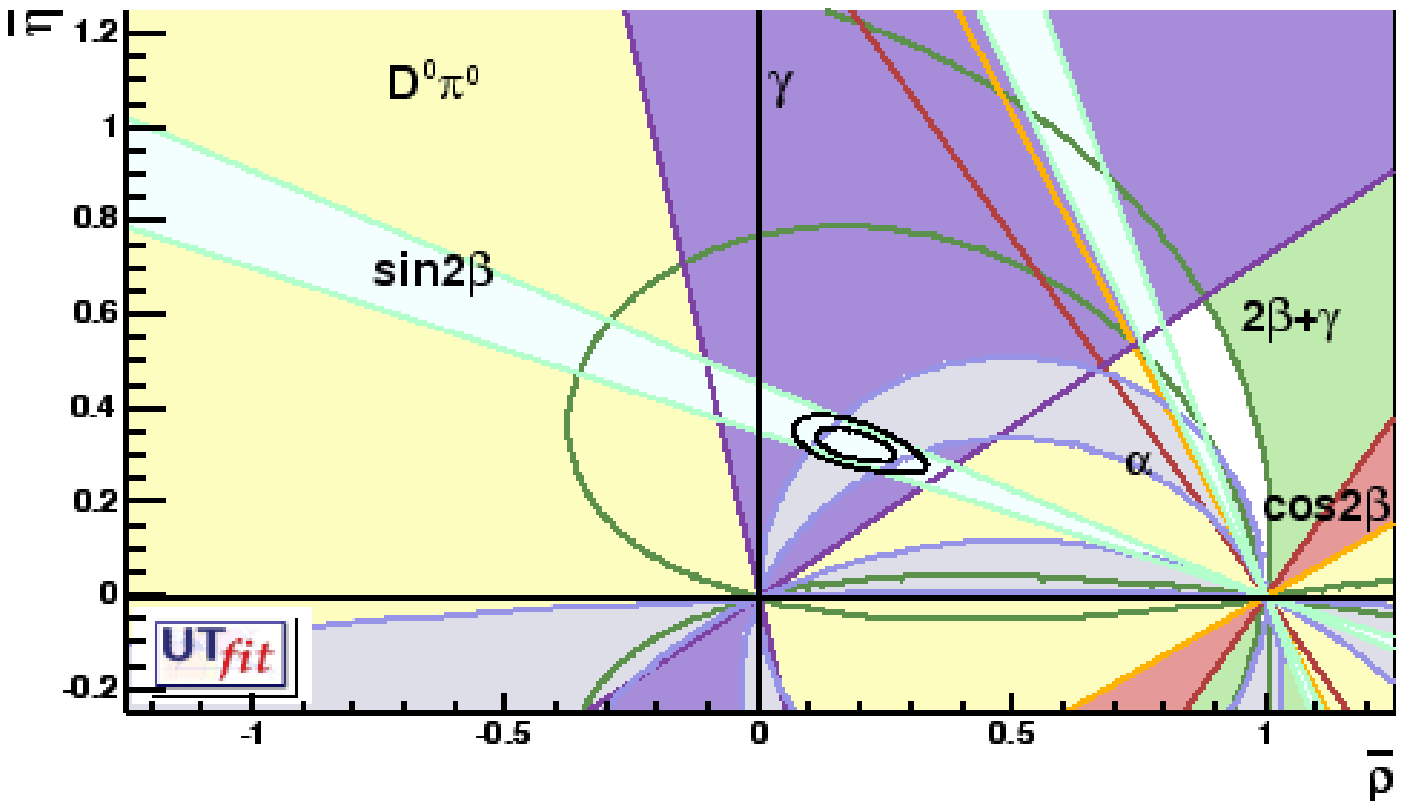}
\epsfxsize 0.49\textwidth
\figurebox{}{}{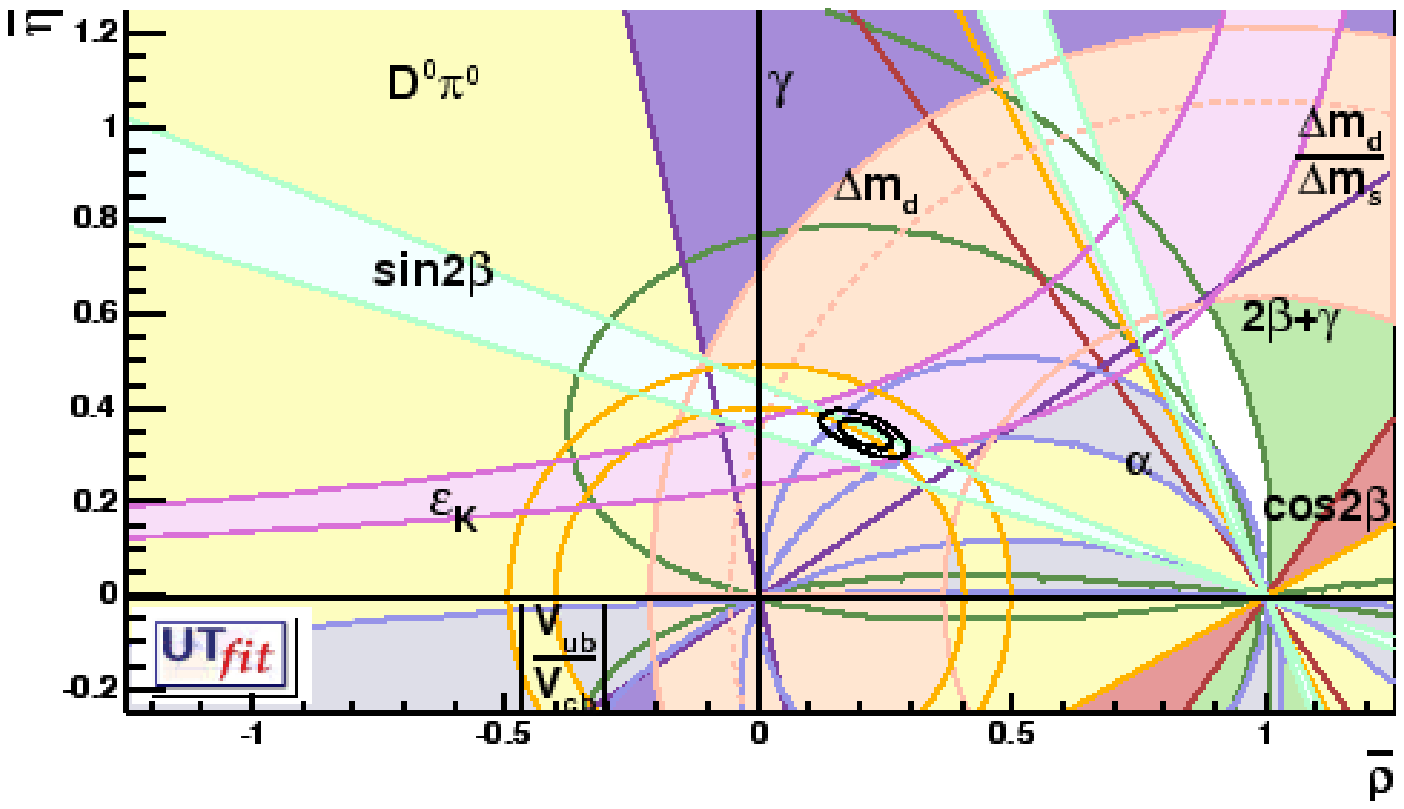}
\caption{The UT obtained without using (top) and using only (center)
  the measurements of the UT angles, and the combined fit result (bottom).}
\label{fig:SMUT}
\end{figure}

A very useful tool to combine the available experimental data in the
quark sector is the Unitarity Triangle (UT) analysis.
\cite{utfit2005,CKMfitter} Thanks to the measurements of
the UT angles recently performed at $B$ factories, which provide a
determination of the UT comparable in accuracy with the one performed
using the other available data, the UT fit is now overconstrained (see
Fig.~\ref{fig:SMUT}). 
It is therefore become possible to add NP contributions to all
quantities entering the UT analysis and to perform a combined fit of
NP contributions and SM parameters.  In general, NP models introduce a
large number of new parameters: flavour changing couplings, short
distance coefficients and matrix elements of new local operators. The
specific list and the actual values of these parameters can only be
determined within a given model. Nevertheless, each of the
meson-antimeson mixing processeses is described by a single amplitude
and can be parameterized, without loss of generality, in terms of two
parameters, which quantify the difference between the full amplitude
snd the SM one.\cite{cfactors} Thus, for instance, in the case of
$B^0_q-\bar{B}^0_q$ mixing we define
\begin{equation} 
  C_{B_q} \, e^{2 i \phi_{B_q}} = \frac{\langle
    B^0_q|H_\mathrm{eff}^\mathrm{full}|\bar{B}^0_q\rangle} {\langle
    B^0_q|H_\mathrm{eff}^\mathrm{SM}|\bar{B}^0_q\rangle}\,, \;
  (q=d,s) 
  \label{eq:paranp} 
\end{equation} 
where
$H_\mathrm{eff}^\mathrm{SM}$ includes only the SM box diagrams, while
$H_\mathrm{eff}^\mathrm{full}$ includes also the NP contributions. As
far as the $K^0-\bar{K}^0$ mixing is concerned, we find it convenient
to introduce a single parameter which relates the imaginary part of
the amplitude to the SM one:
\begin{equation} 
  C_{\epsilon_K} = \frac{\mathrm{Im}[\langle
    K^0|H_{\mathrm{eff}}^{\mathrm{full}}|\bar{K}^0\rangle]}
  {\mathrm{Im}[\langle
    K^0|H_{\mathrm{eff}}^{\mathrm{SM}}|\bar{K}^0\rangle]}\,.
  \label{eq:ceps} 
\end{equation} 
Therefore, all NP effects in $\Delta F=2$ transitions are
parameterized in terms of three real quantities, $C_{B_d}$,
$\phi_{B_d}$ and $C_{\epsilon_K}$.  NP in the $B_s$ sector is not
considered, due to the lack of experimental information, since both
$\Delta m_s$ and $A_\mathrm{CP}(B_s \to J/\psi \phi)$ are not yet
measured.

\begin{figure*}[p]
  \begin{tabular}{cc}
    \epsfxsize 0.49\textwidth
    \figurebox{}{}{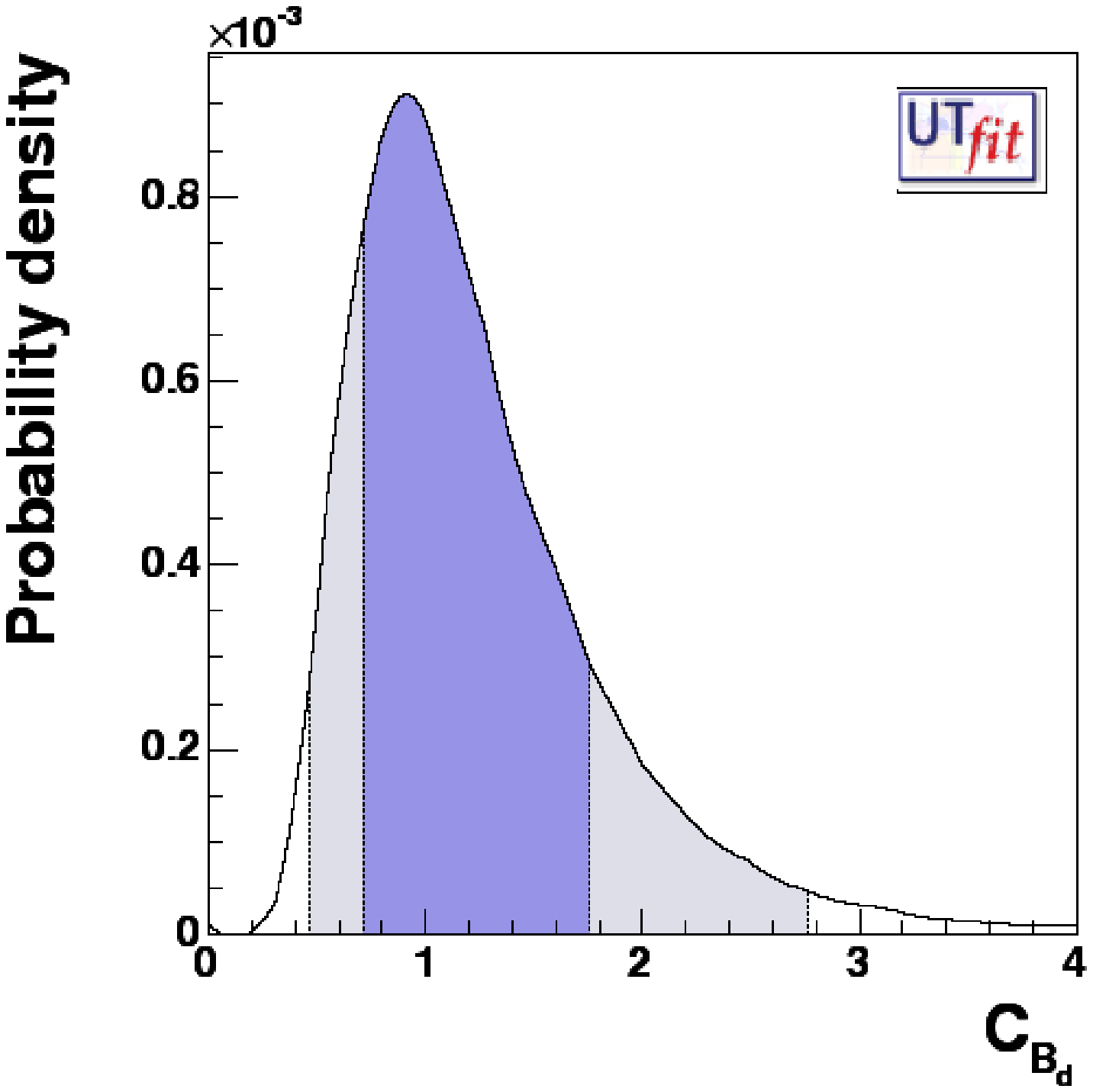} &
    \epsfxsize 0.49\textwidth
    \figurebox{}{}{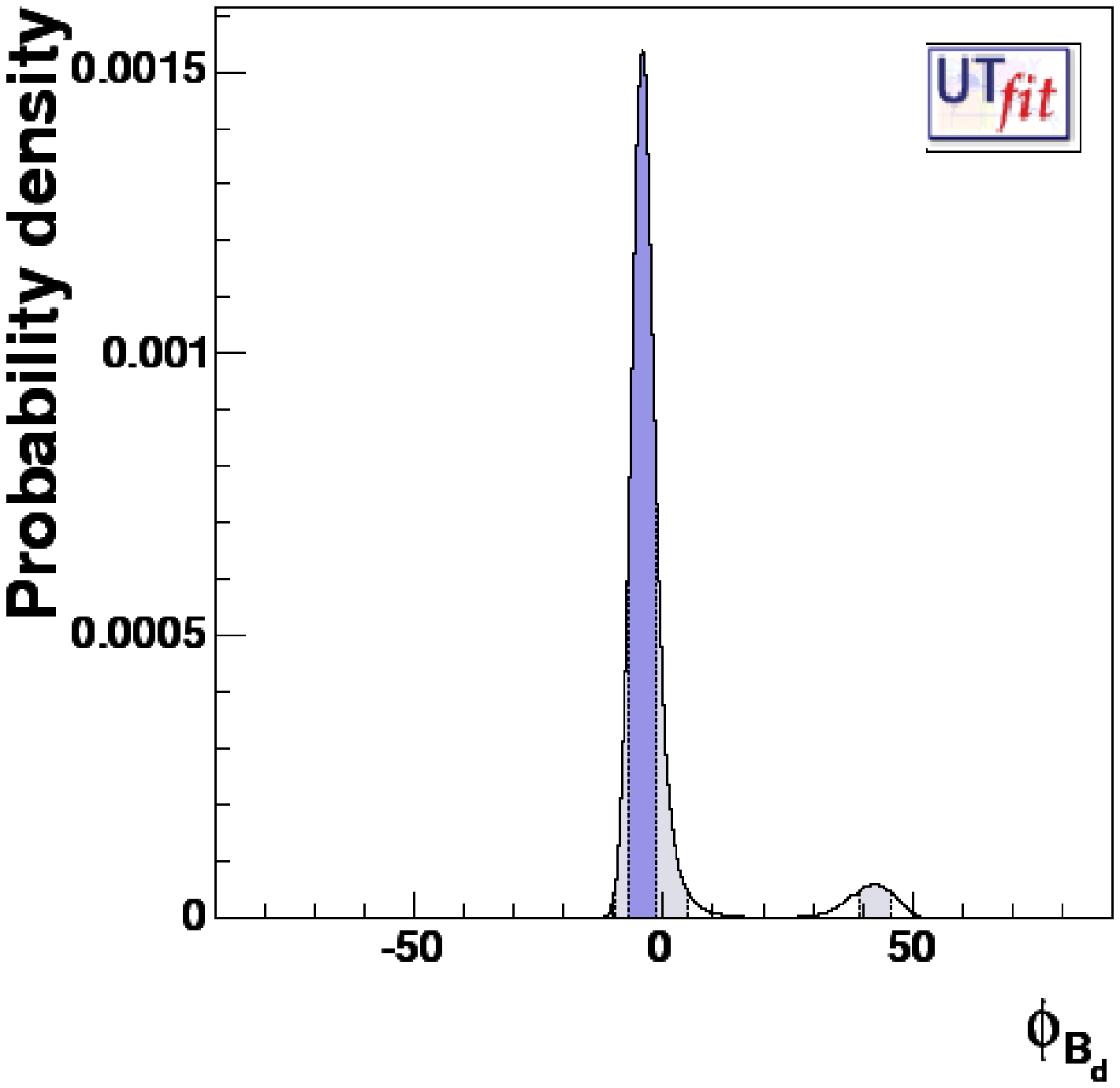} \\
    \epsfxsize 0.49\textwidth
    \figurebox{}{}{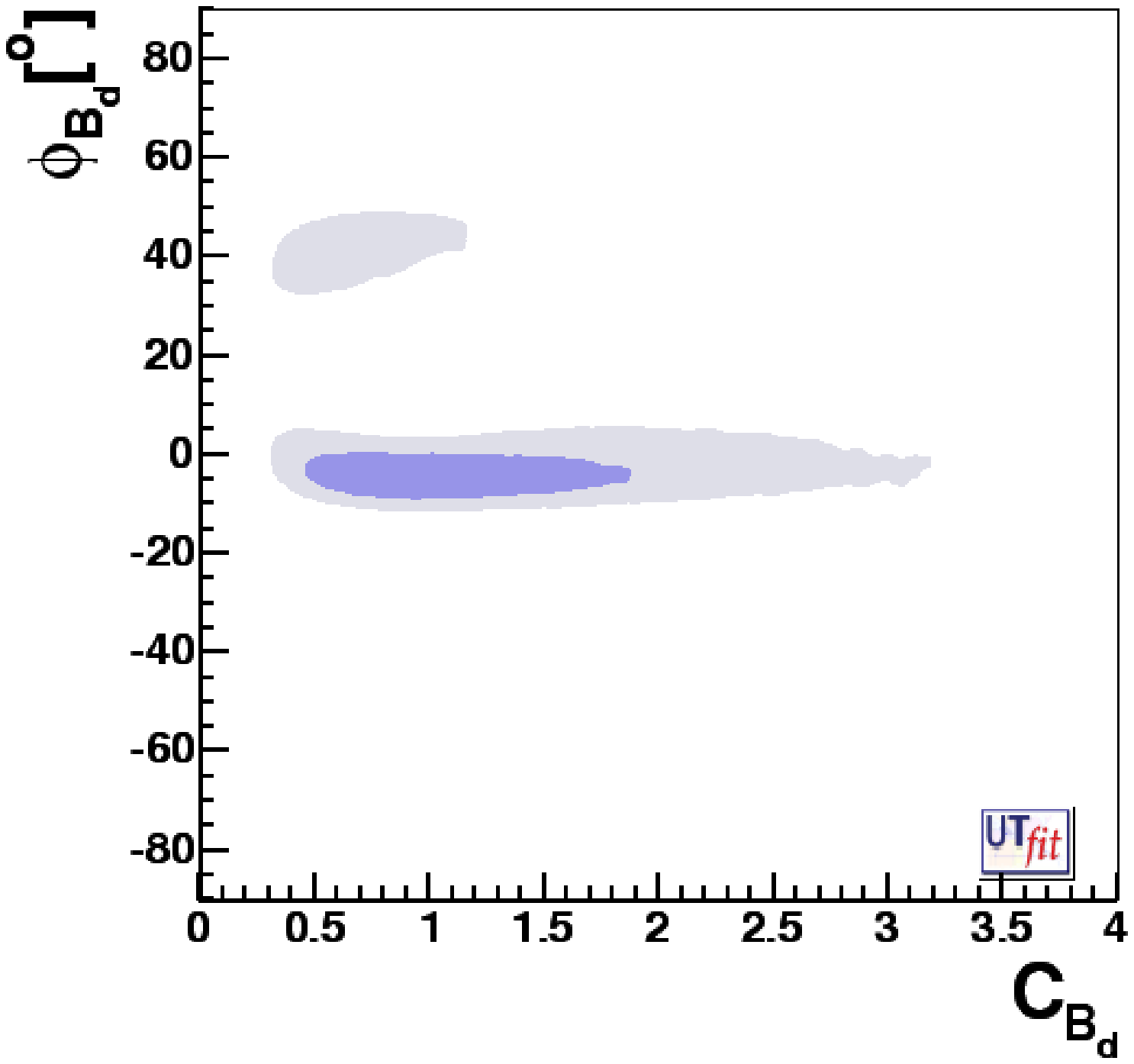}&
    \epsfxsize 0.49\textwidth
    \figurebox{}{}{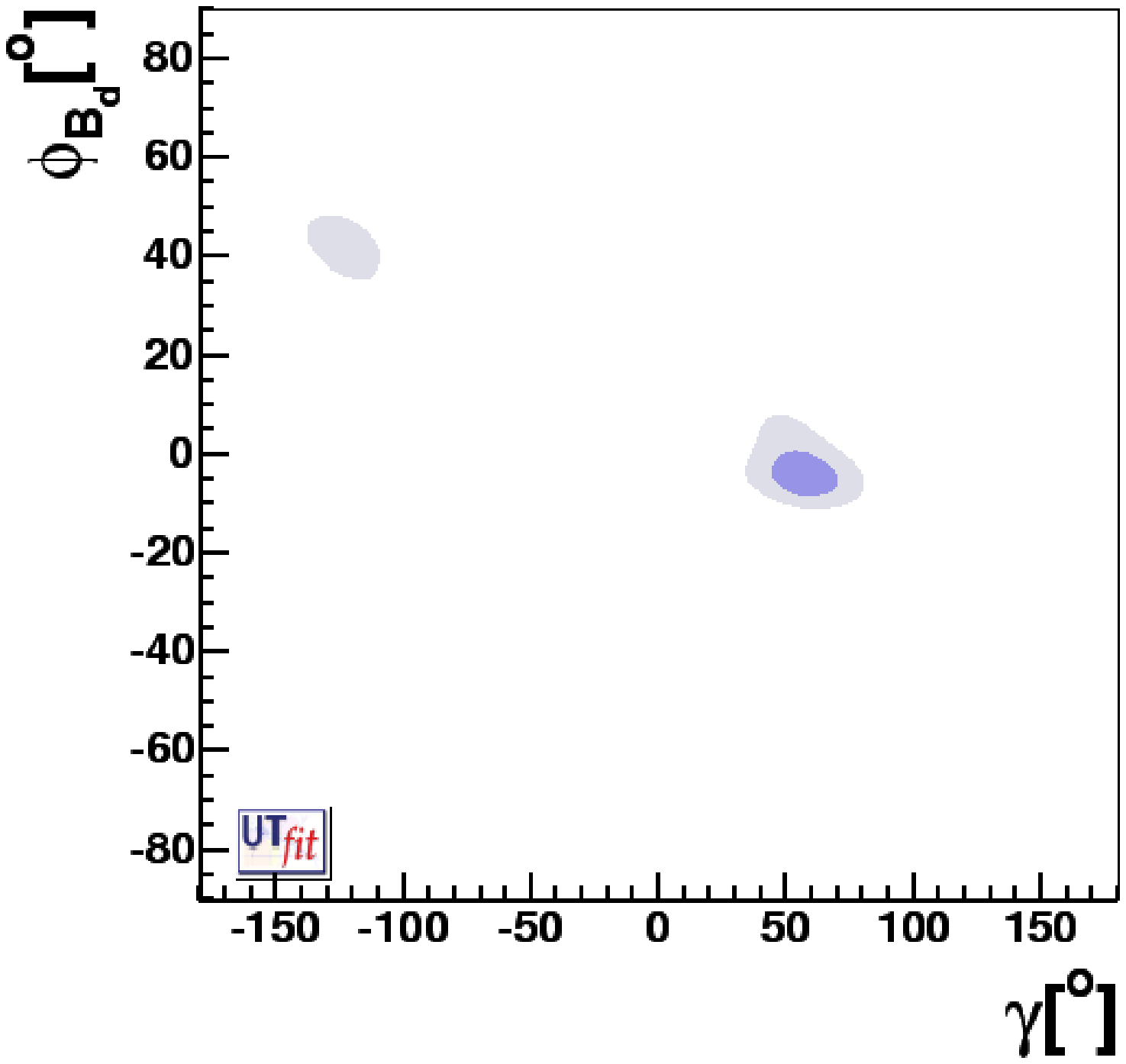} \\
    \epsfxsize 0.49\textwidth
    \figurebox{}{}{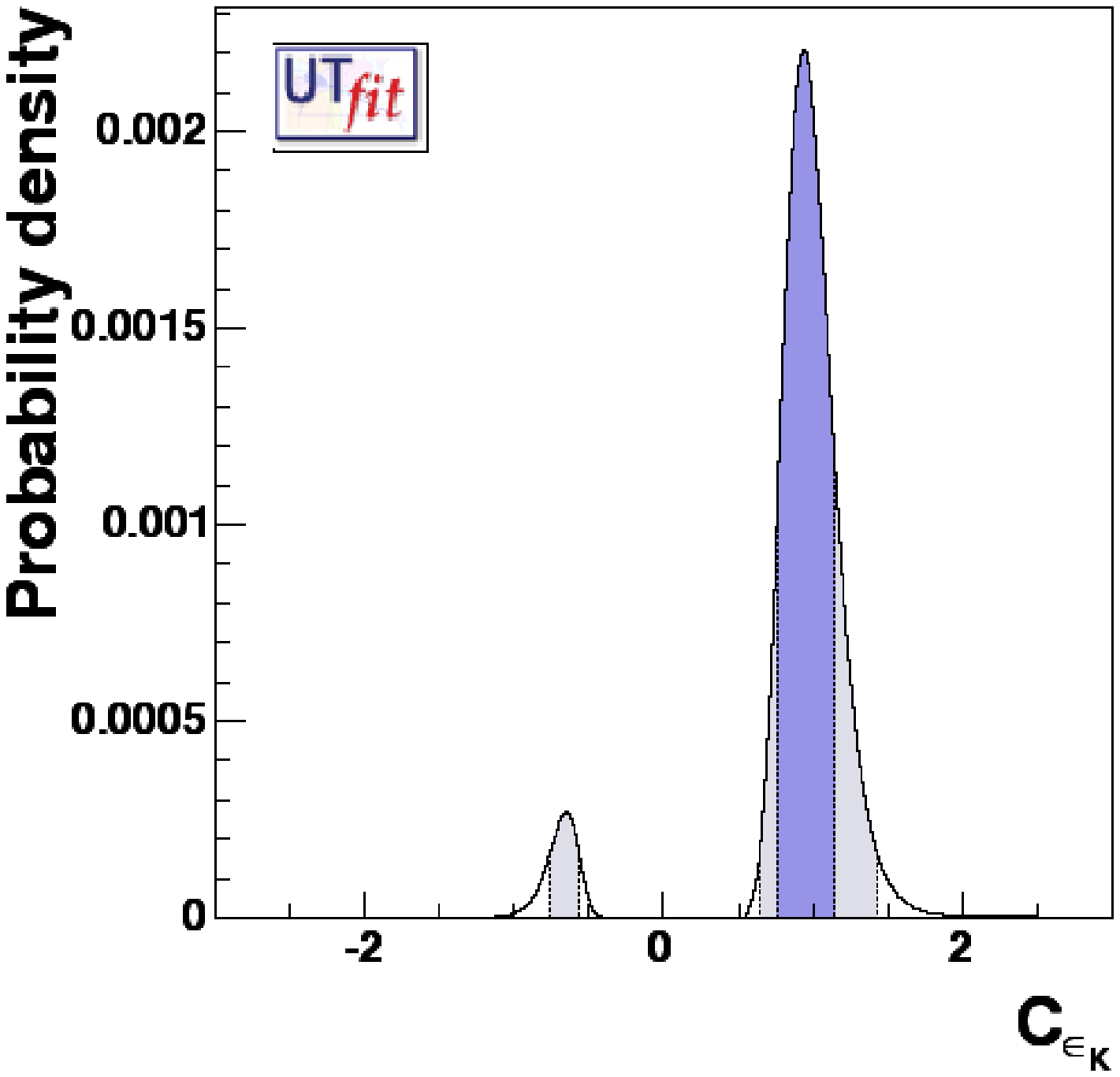} &
    \epsfxsize 0.49\textwidth
    \figurebox{}{}{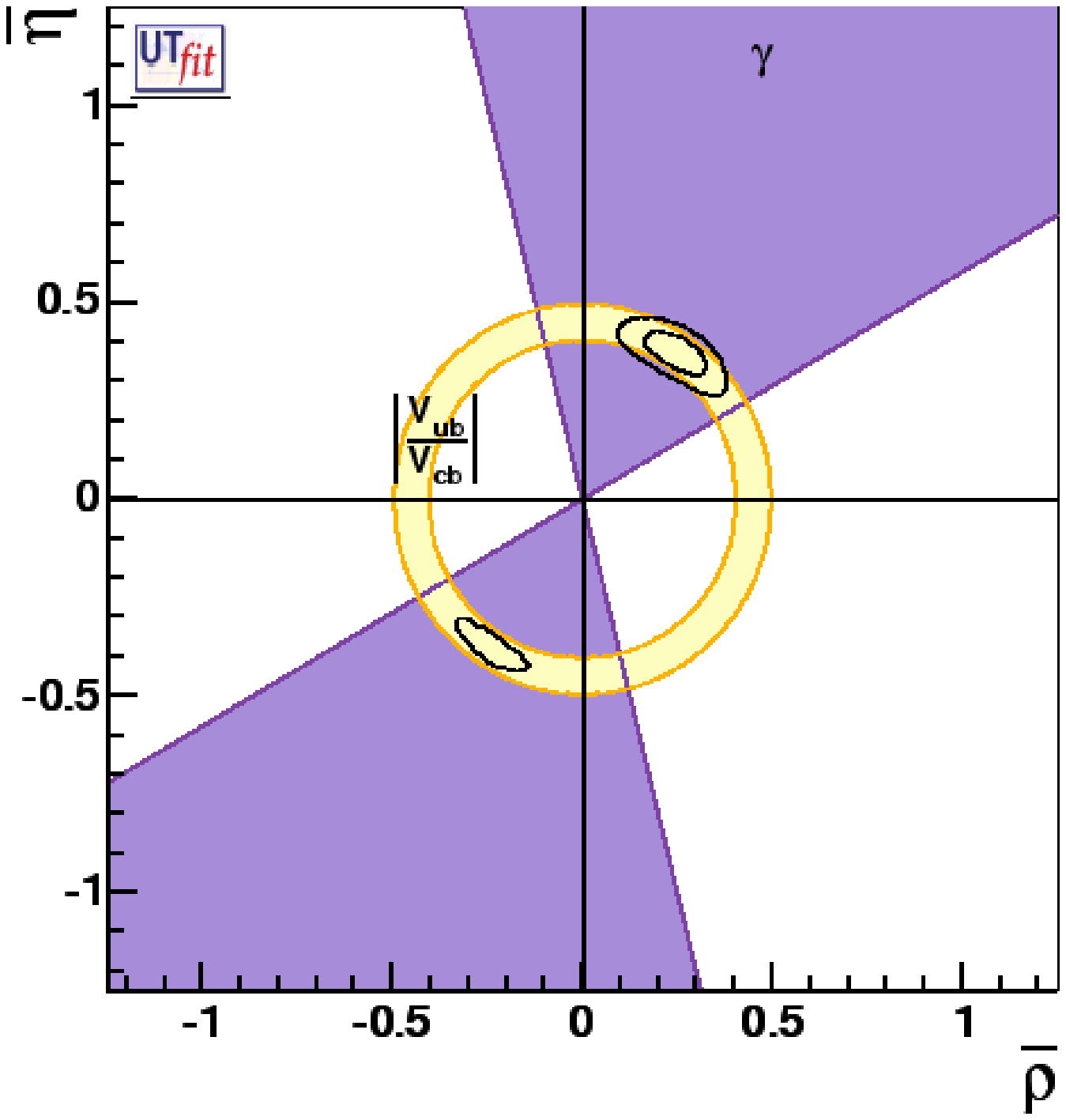}
\end{tabular}
  \caption{From top to bottom and from left to right, p.d.f.'s for
    $C_{B_d}$, $\phi_{B_d}$, $\phi_{B_d}\,vs.\,C_{B_d}$,
    $\phi_{B_d}\,vs.\,\gamma$, $C_{\epsilon_K}$ and the selected
    region on the $\bar\rho-\bar\eta$ plane obtained from the NP
    analysis. In the last plot, selected regions
    corresponding to $68\%$ and $95\%$ probability are shown, together
    with $95\%$ probability regions for $\gamma$ (from $DK$ final
    states) and $\vert V_{ub}/V_{cb}\vert$.  Dark (light) areas
    correspond to the $68\%$ ($95\%$) probability region.}
  \label{fig:NP}
\end{figure*}

NP effects in $\Delta B=1$ transitions can also affect some of the
measurements entering the UT analysis, in particular the measurements
of $\alpha$ and $A_\mathrm{SL}$.\cite{utfitNP} However, under the
hypothesis that NP contributions are mainly $\Delta I=1/2$, their effect
can be taken into account in the fit of the $B\to \pi
\pi,\rho \pi, \rho \rho$ decay amplitudes. Concerning $A_\mathrm{SL}$,
penguins only enter at the Next-to-Leading order and therefore NP in
$\Delta B=1$ transitions produces subdominant effects with respect to
the leading $\Delta B=2$ contribution.

The results obtained in a global fit for $C_{B_d}$, $C_{\epsilon_K}$,
$C_{B_d}$ vs. $\phi_{B_d}$, and $\gamma$ vs. $\phi_{B_d}$ are shown in
Fig.~\ref{fig:NP}, together with the corresponding regions in the
$\bar\rho$--$\bar\eta$ plane.\cite{utfitNP}

\begin{figure}[h!]
  \begin{tabular}{cc}
    \epsfxsize 0.22\textwidth
    \figurebox{}{}{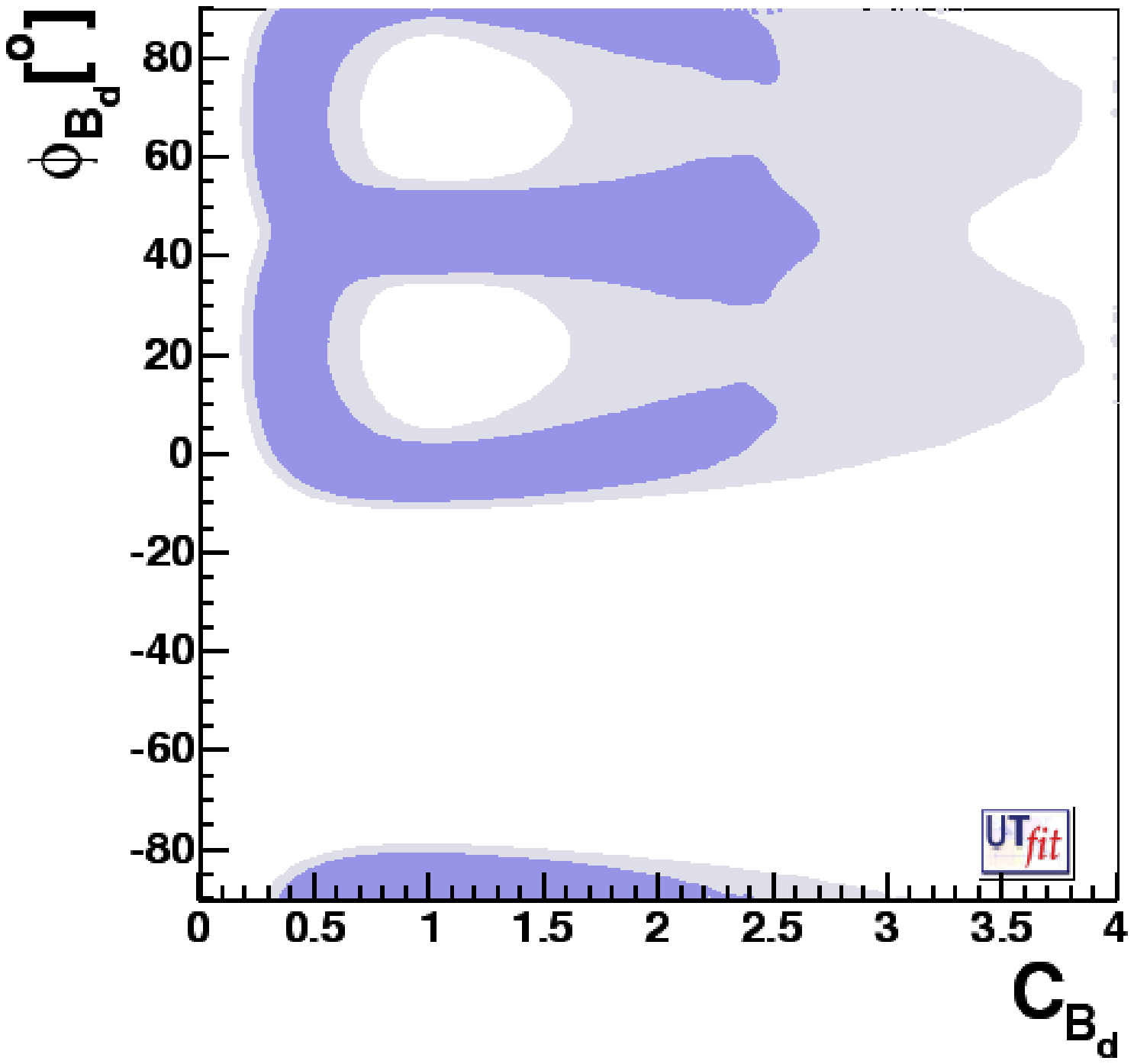} &
    \epsfxsize 0.22\textwidth
    \figurebox{}{}{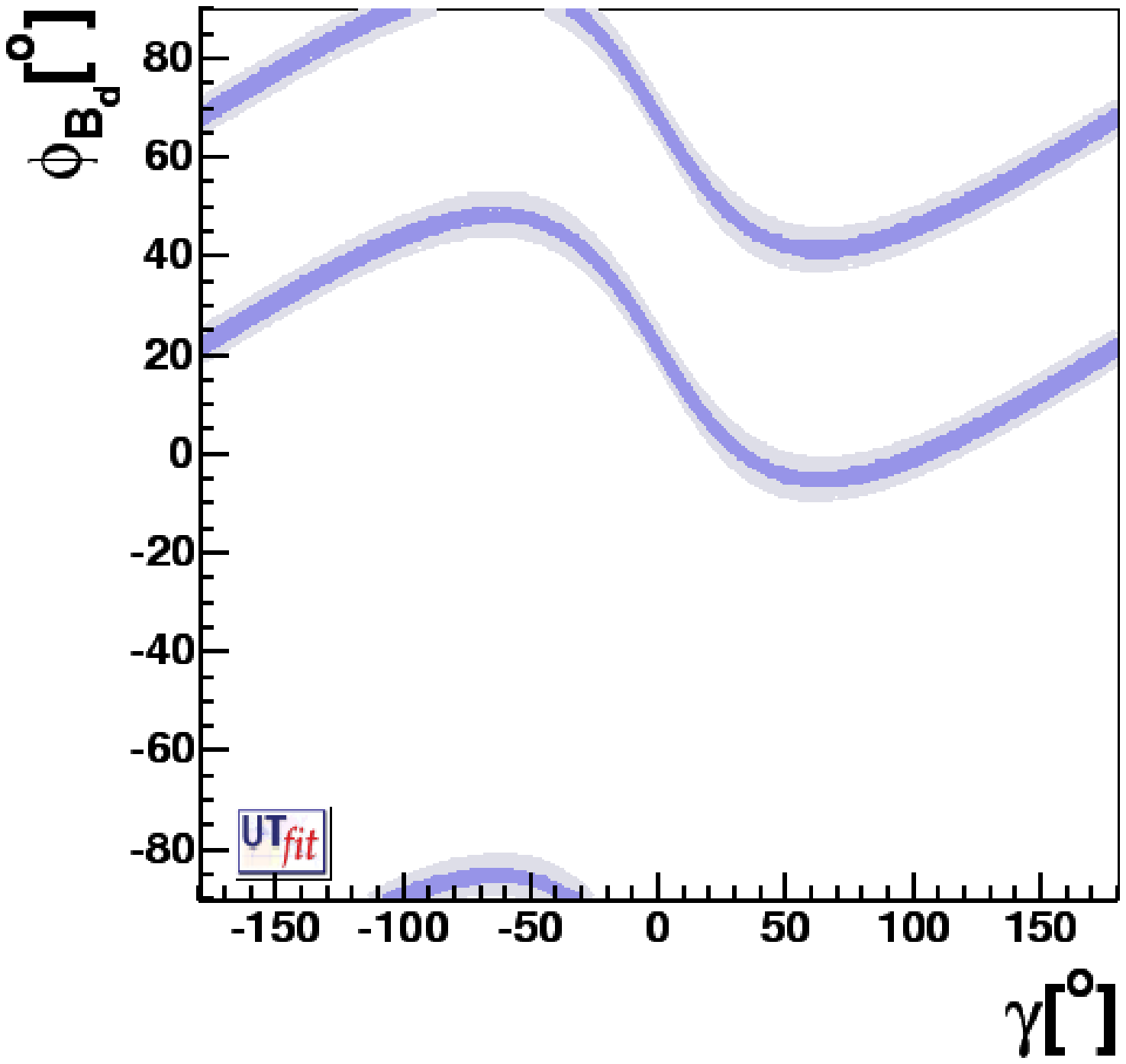} \\
    \epsfxsize 0.22\textwidth
    \figurebox{}{}{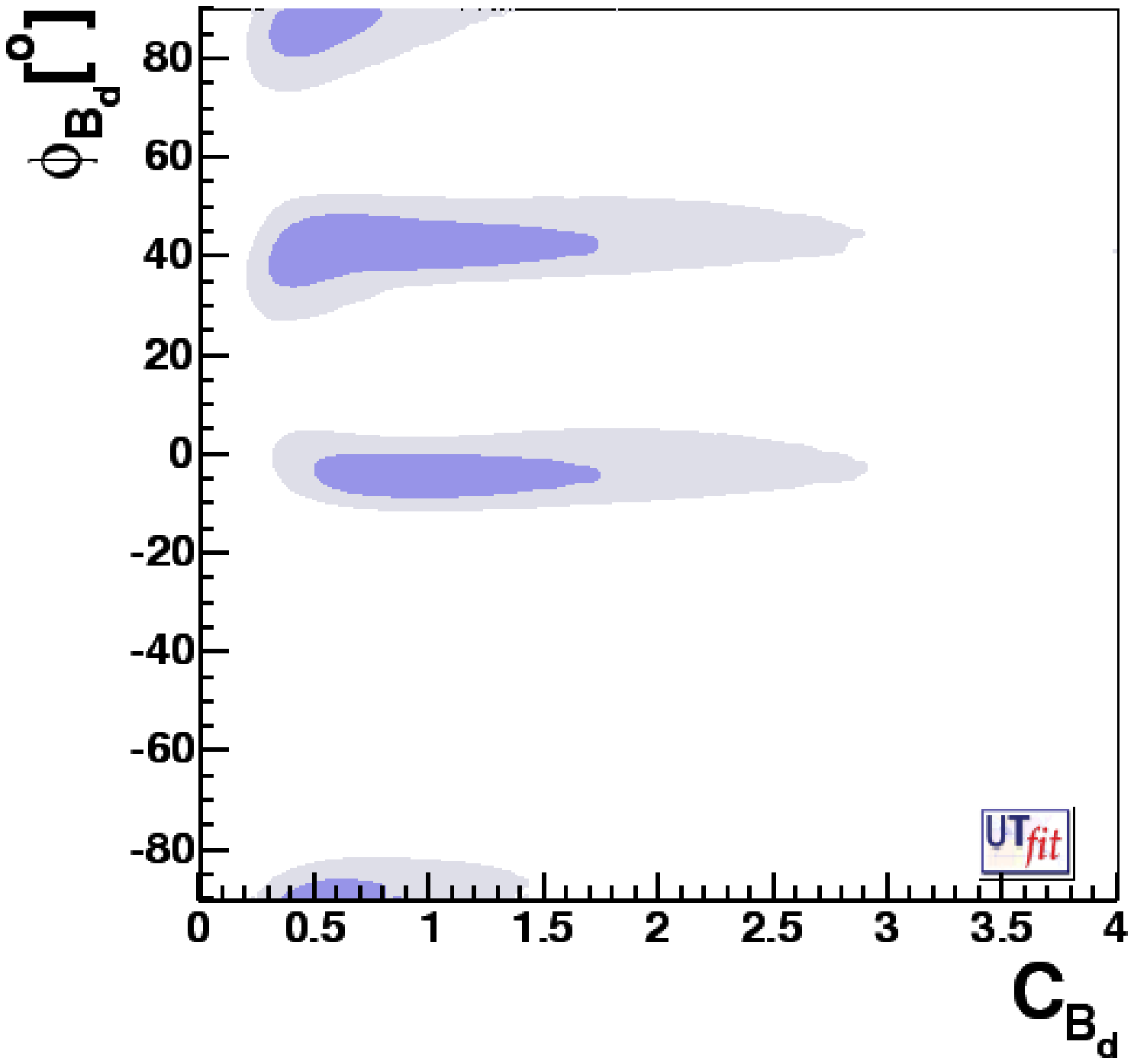} &
    \epsfxsize 0.22\textwidth
    \figurebox{}{}{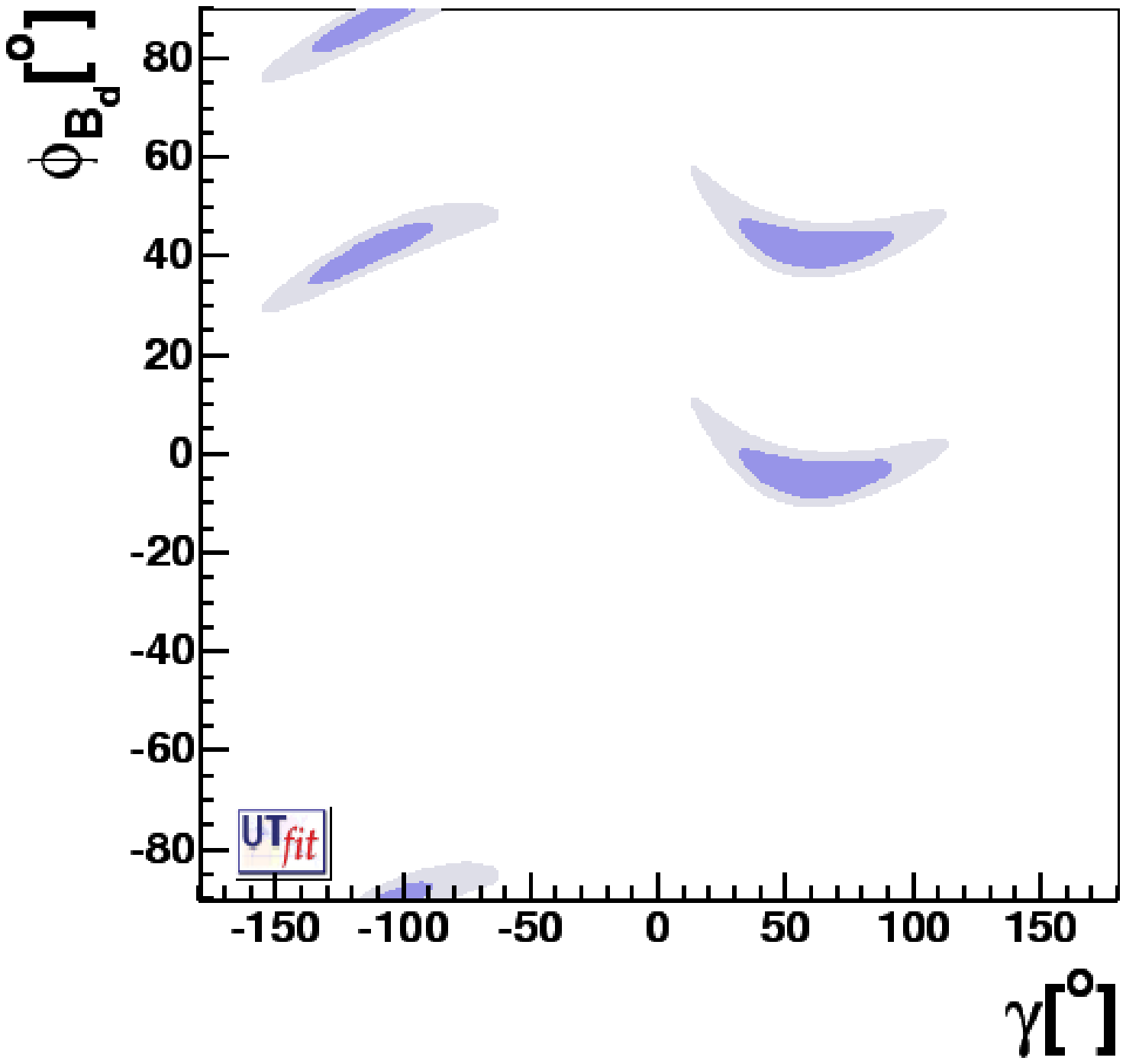} \\
    \epsfxsize 0.22\textwidth
    \figurebox{}{}{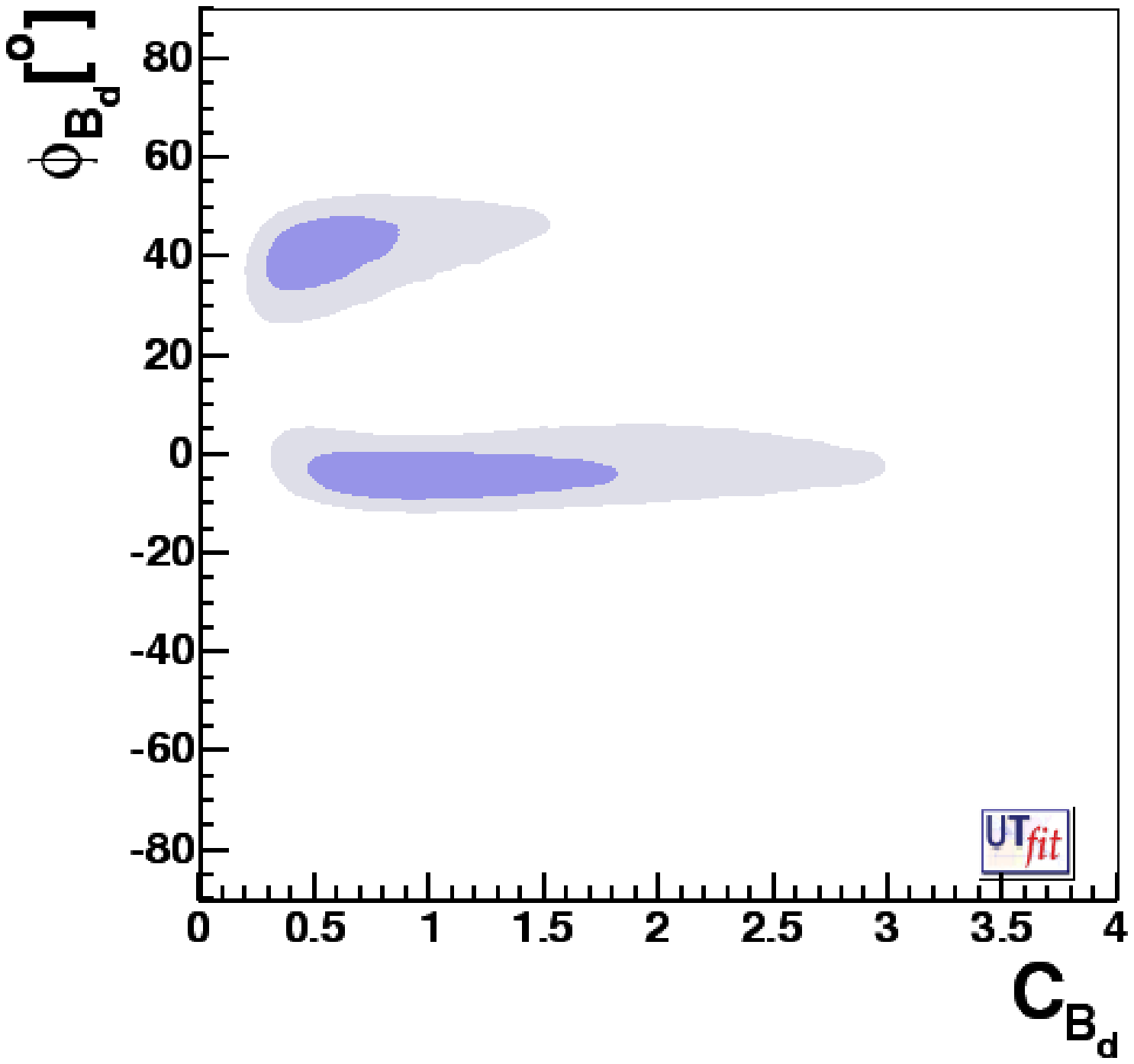} &
    \epsfxsize 0.22\textwidth
    \figurebox{}{}{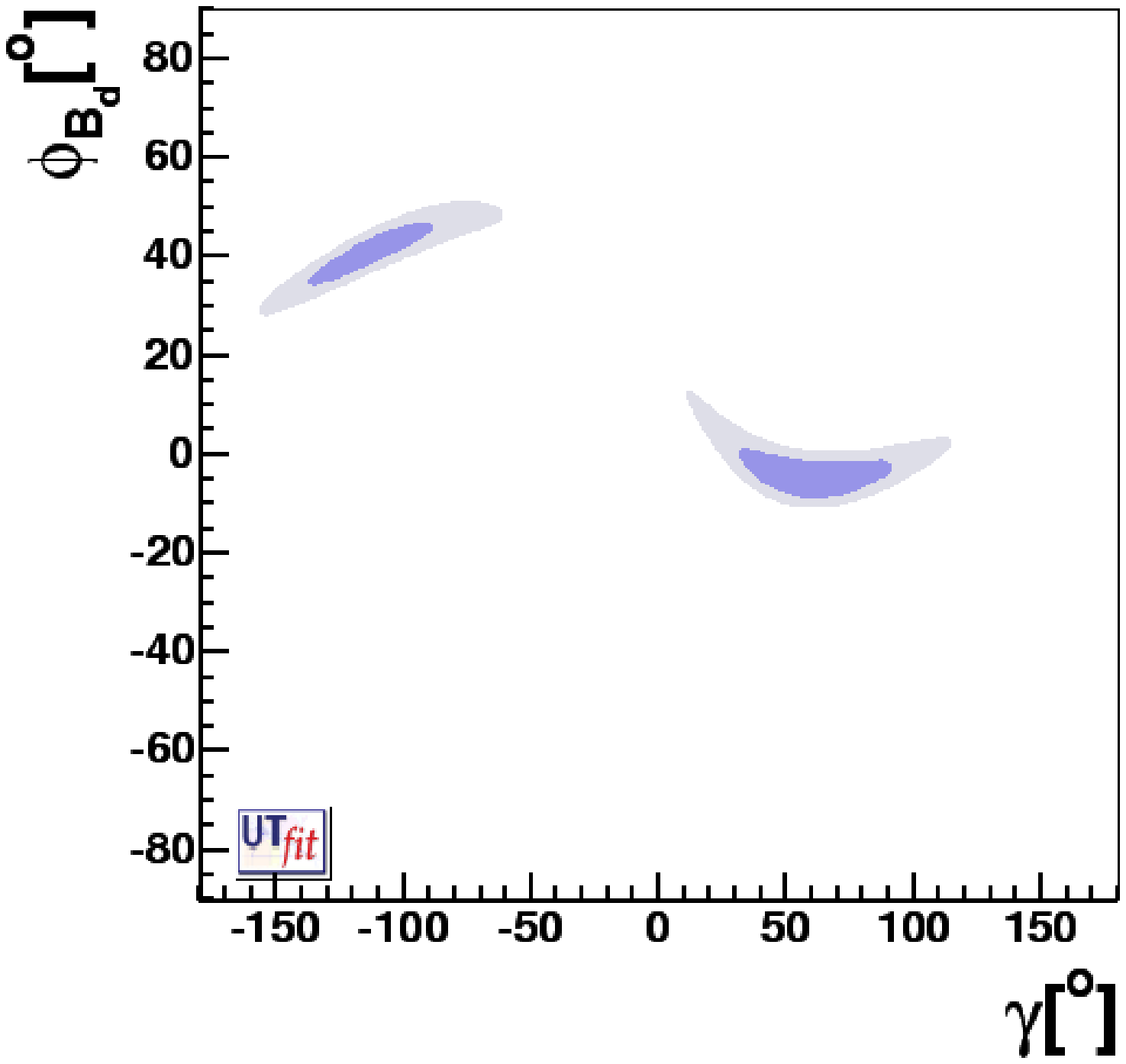} \\
    \epsfxsize 0.22\textwidth
    \figurebox{}{}{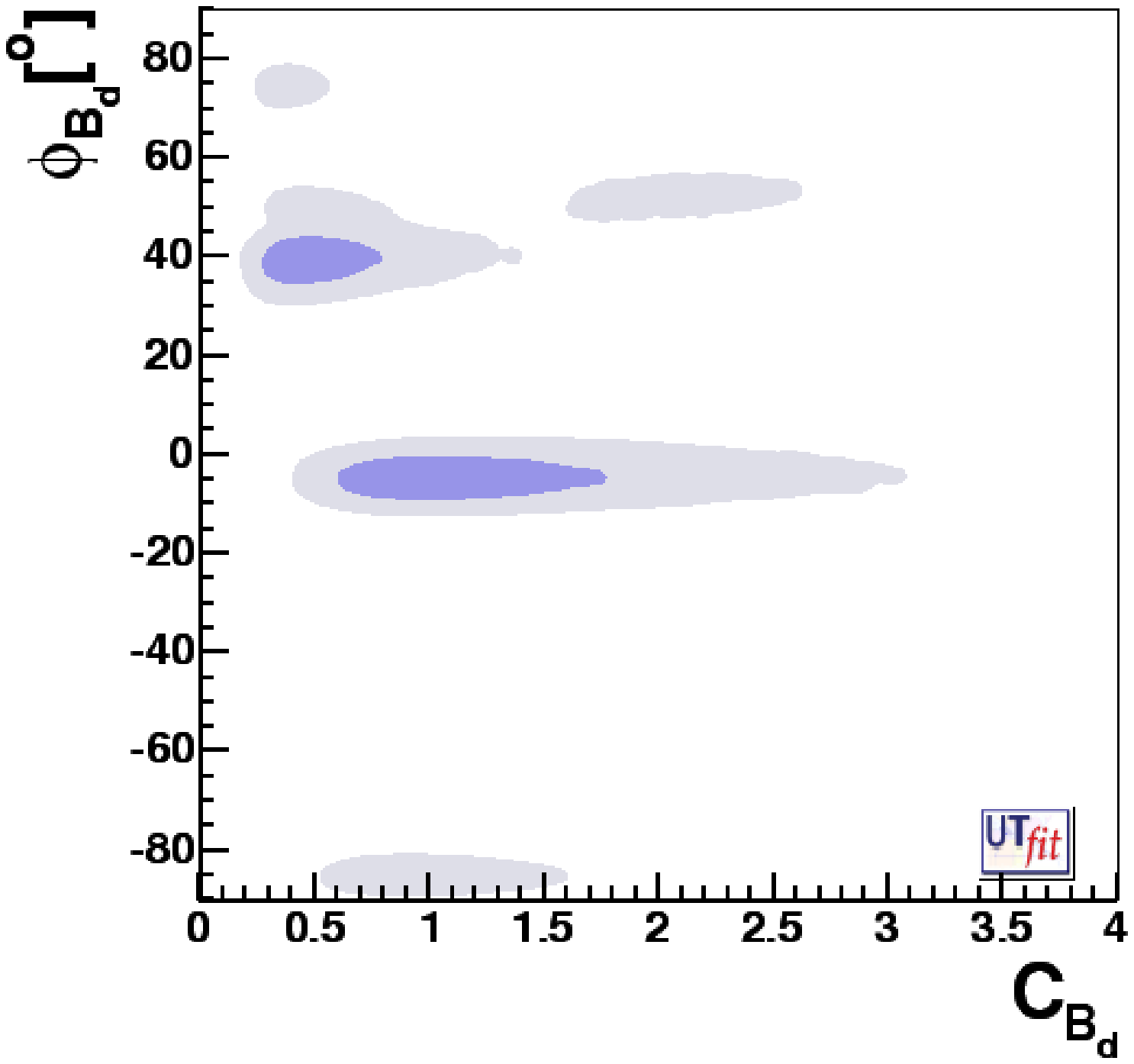} &
    \epsfxsize 0.22\textwidth
    \figurebox{}{}{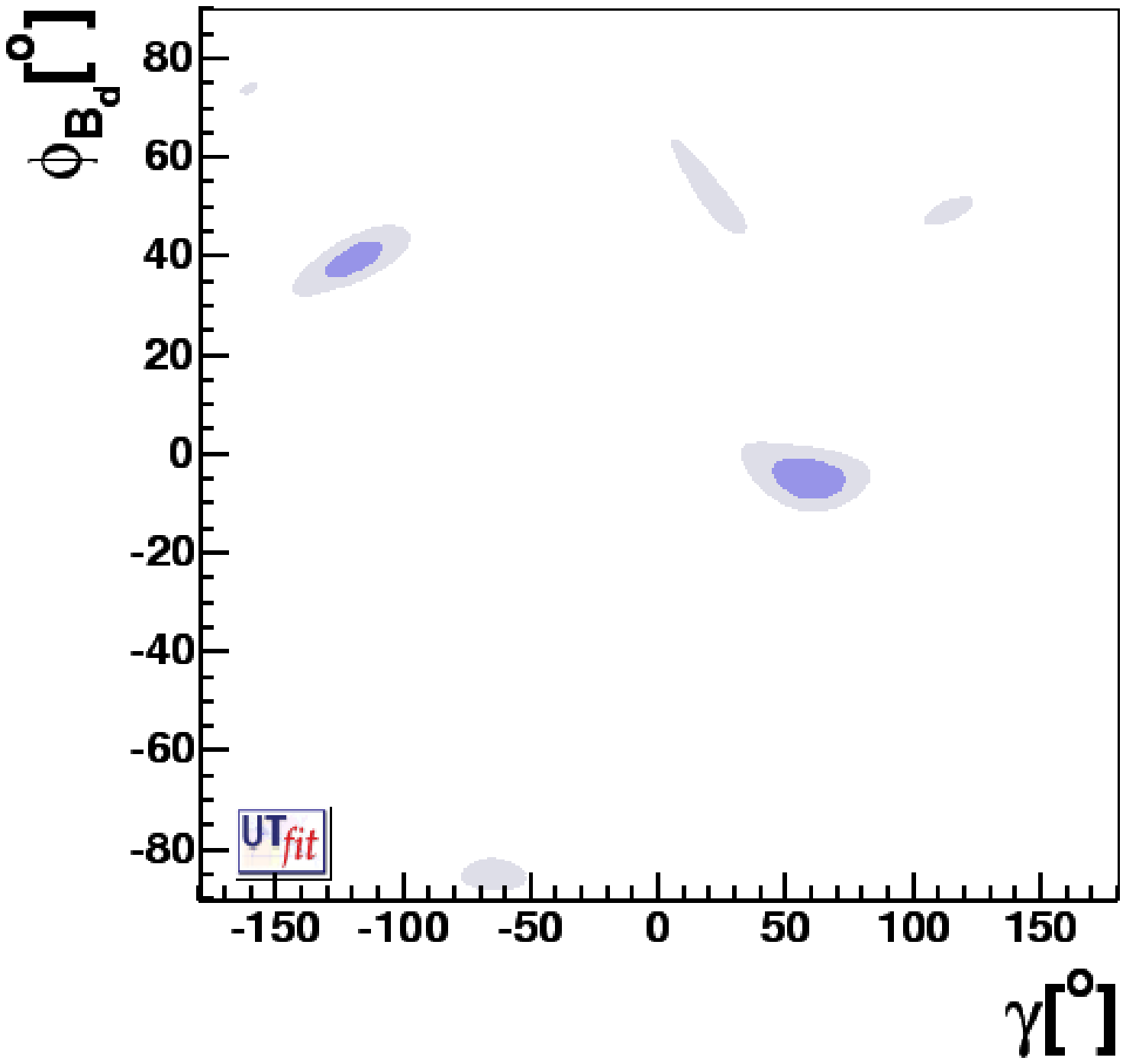} 
  \end{tabular}
  \caption{From top to bottom: distributions of
    $\phi_{B_d}$ vs. $C_{B_d}$ (left) and $\phi_{B_d}$vs. $\gamma$
    (right) using the following constraints: i) $|V_{ub}/V_{cb}|$,
    $\Delta m_d$, $\varepsilon_K$ and $\sin 2 \beta$; ii)
    the constraints in i) plus $\gamma$; iii) the
    constraints in ii) plus $\cos 2 \beta$ from $B_d \to J/\psi K^*$
    and $\beta$ from $B \to D h^0$; iv) the constraints in
    ii) plus $\alpha$.}
  \label{fig:individual}
\end{figure}

To illustrate the impact of the various constraints on the analysis,
in Fig.~\ref{fig:individual} we show the selected regions in the
$\phi_{B_d}$ vs. $C_{B_d}$ and $\phi_{B_d}$ vs. $\gamma$ planes using
different combinations of constraints. The first row represents the
pre-2004 situation, when only $|V_{ub}/V_{cb}|$, $\Delta m_d$,
$\varepsilon_K$ and $\sin 2 \beta$ were available, selecting a
continuous band for $\phi_{B_d}$ as a function of $\gamma$ and a broad
region for $C_{B_d}$. Adding the determination of $\gamma$ (second
row), only four regions in the $\phi_{B_d}$ vs. $\gamma$ plane
survive, two of which overlap in the $\phi_{B_d}$ vs. $C_{B_d}$ plane.
Two of these solutions have values of $\cos 2 (\beta + \phi_{B_d})$
and $\alpha - \phi_{B_d}$ different from the SM predictions, and are
therefore disfavoured by $(\cos 2 \beta)^\mathrm{exp}$ and by the
measurement of $(2 \beta)^\mathrm{exp}$ from $B \to D h^0$ decays, and
by $\alpha^\mathrm{exp}$ (third and fourth row respectively). On the
other hand, the remaining solution has a very large value for
$A_\mathrm{SL}$ and is therefore disfavoured by
$A_\mathrm{SL}^\mathrm{exp}$, leading to the final results already
presented in Fig.~\ref{fig:NP}. The numerical results of the analysis
can be found in ref.~\cite{utfitNP} (see
ref.~\cite{CKMfitter,previous} for previous analyses).

\begin{figure}[h!]
  \epsfxsize 0.45\textwidth
  \figurebox{}{}{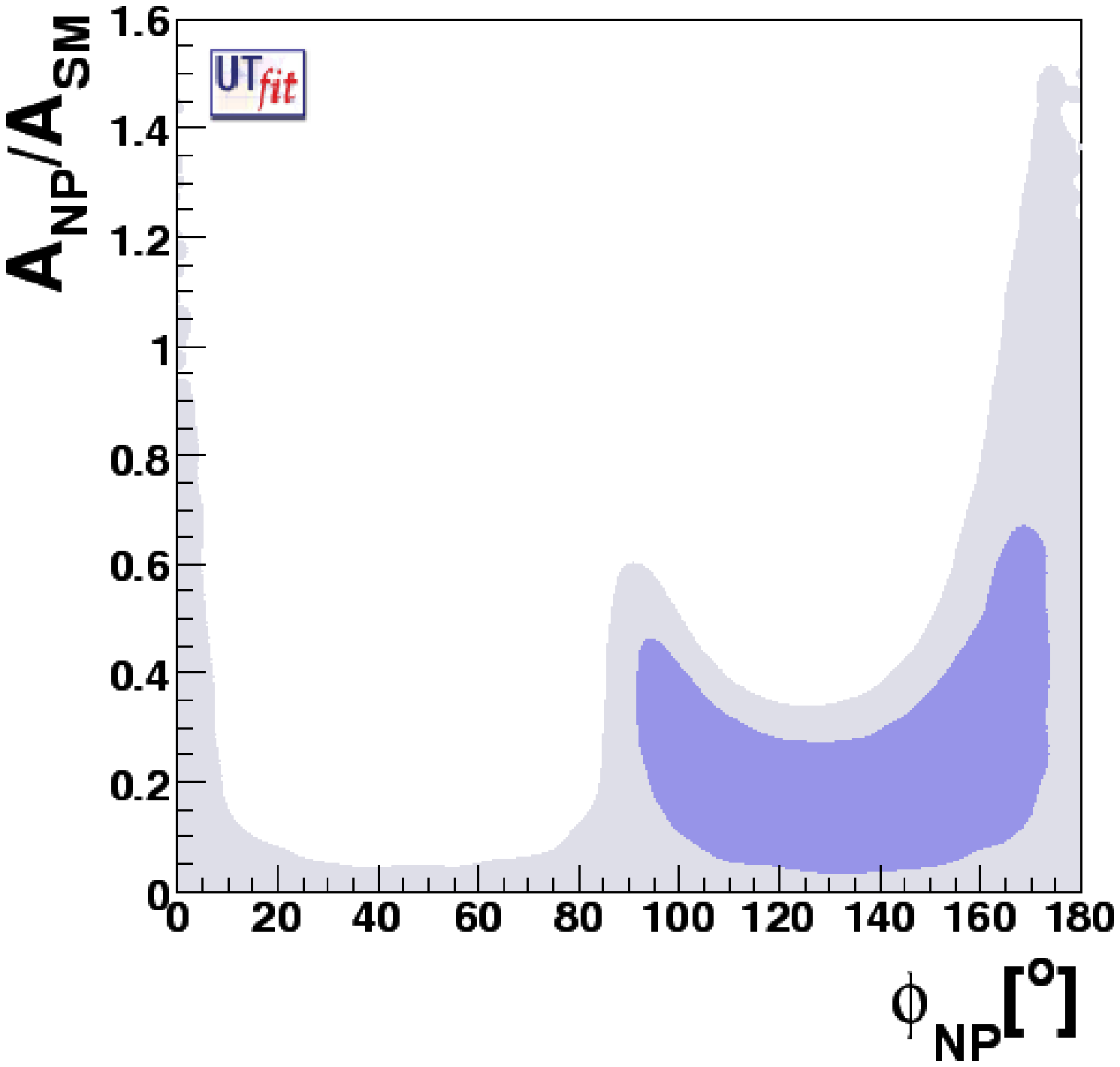}
\caption{P.d.f. in the $(A_\mathrm{NP}/A_\mathrm{SM})$ vs.
    $\phi_\mathrm{NP}$ plane for NP in the $|\Delta B|=2$ sector (see
    Eq.~(\ref{eq:asmanp})).}
\label{fig:achilleplot}
\end{figure}

Before concluding this section, let us analyze more in detail the
results in Fig.~\ref{fig:NP}. Writing
\begin{equation}
  \label{eq:asmanp}
  C_{B_d}e^{2 i \phi_{B_d}} = \frac{A_\mathrm{SM} e^{2 i \beta} +
    A_\mathrm{NP} e^{2 i (\beta + \phi_\mathrm{NP})}}{A_\mathrm{SM}
    e^{2 i \beta}}\,, 
\end{equation}
and given the p.d.f. for $C_{B_d}$ and $\phi_{B_d}$, we can derive the
p.d.f. in the $(A_\mathrm{NP}/A_\mathrm{SM})$ vs. $\phi_\mathrm{NP}$
plane. The result is reported in Fig.~\ref{fig:achilleplot}. We see
that the NP contribution can be substantial if its phase is close to
the SM phase, while for arbitrary phases its magnitude has to be much
smaller than the SM one. Notice that, with the latest data, the SM
($\phi_{B_d}=0$) is disfavoured at $68\%$ probability due to a slight
disagreement between $\sin 2\beta$ and $|V_{ub}/V_{cb}|$.  This
requires $A_\mathrm{NP}\neq 0$ and $\phi_\mathrm{NP}\neq 0$. For the
same reason, $\phi_\mathrm{NP}>90^\circ$ at $68\%$ probability and the
plot is not symmetric around $\phi_\mathrm{NP}=90^\circ$. 

Assuming that the small but non-vanishing value for $\phi_{B_d}$ we
obtained is just due to a statistical fluctuation, the result of our
analysis points either towards models with no new source of flavour
and CP violation beyond the ones present in the SM (Minimal Flavour
Violation, MFV), or towards models in which new sources of flavour and
CP violation are only present in $b \to s$ transitions. In the rest of
this talk we will consider these two possibilities, starting from the
former.

\section{MFV models}
\label{sec:uut}

We now specialize to the case of MFV. Making the basic assumption that
the only source of flavour and CP violation is in the Yukawa
couplings,\cite{gino} it can be shown that the phase of $|\Delta B|=2$
amplitudes is unaffected by NP, and so is the ratio $\Delta m_s/\Delta
m_d$. This allows the determination of the Universal Unitarity
Triangle independent on NP effects, based on $\vert
V_{ub}/V_{cb}\vert$, $\gamma$, $A_{CP}(B \to J/\Psi K^{(*)})$, $\beta$
from $B \to D^0h^0$, $\alpha$, and $\Delta m_s/\Delta m_d$.\cite{uut}
We present here the determination of the UUT, which is independent of
NP contributions in the context of MFV models.  The details of the
analysis and the upper bounds on NP contributions that can be derived
from it can be found in ref.~\cite{utfitNP}

\begin{figure}[ht!]
  \epsfxsize 0.49\textwidth
  \figurebox{}{}{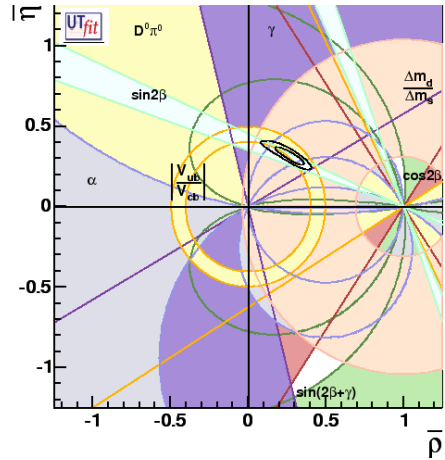}
\caption{The selected region on $\bar\rho$-$\bar\eta$ plane obtained from 
the determination of the UUT.}
\label{fig:uut}
\end{figure}

In Fig.~\ref{fig:uut} we show the allowed region in the
$\bar\rho-\bar\eta$ plane for the UUT. The corresponding values
and ranges are reported in Tab.~\ref{tab:uut}. The most important
differences with respect to the general case are that i) the lower
bound on $\Delta m_s$ forbids the solution in the third quadrant, and ii)
the constraint from $\sin 2 \beta$ is now effective, so that we are
left with a region very similar to the SM one. 

\begin{table}[ht!]
  \caption{Results of the UUT analysis.}
  \label{tab:uut} 
  \begin{tabular}{ccc}
    \hline
    & UUT ($68\%$)  & UUT ($95\%$) \\
    \hline
    $\bar\rho$                   & 0.259 $\pm$ 0.068 & $ [0.107,\,0.376]$ \\
    $\bar\eta$                   & 0.320 $\pm$ 0.042 & $ [0.241,\,0.399]$ \\
    $\sin 2\beta$                      & 0.728 $\pm$ 0.031 & $ [0.668,\,0.778]$ \\
    $\alpha[^{\circ}]$          & 105  $\pm$ 11   & $ [81,\,124]$  \\
    $\gamma[^{\circ}]$          & 51  $\pm$ 10   & $ [33,\,75]$   \\
    $[2\beta+\gamma][^{\circ}]$ & 98 $\pm$ 12   & $ [77,\,123]$  \\
    $\Delta m_s$ [ps$^{-1}$] & 20.6 $\pm$ 5.6 & $ [10.6,\,32.6]$  \\
    \hline 
  \end{tabular}
\end{table}

Starting from the determination of the UUT, one can study rare decays
in MFV models.\cite{BurasMFV} In general, a model-independent analysis
of rare decays is complicated by the large number of higher
dimensional operators that can contribute beyond the
SM.\cite{hilleretal} The situation drastically simplifies in MFV
models, where (excluding large $\tan \beta$ scenarios) no new
operators arise beyond those generated by $W$ exchange. Since the mass
scale of NP must be higher than $M_W$, we can further restrict our
attention to operators up to dimension five, since higher dimensional
operators will suffer a stronger suppression by the scale of NP. In
this way, we are left with NP contributions to two operators only: the
FCNC $Z$ and magnetic vertices.\footnote{The chromomagnetic vertex
  should also be considered, but this is not necessary for the
  analysis presented here.\cite{BurasMFV}} NP contributions can be
reabsorbed in a redefinition of the SM coefficients of these
operators: $C = C_\mathrm{SM} + \Delta C$ for the $Z$ vertex and
$C_7^\mathrm{eff} = C_{7\mathrm{SM}}^\mathrm{eff} + \Delta
C_7^\mathrm{eff}$ for the magnetic operator.\footnote{We find it
  convenient to redefine the $C$ function at the electroweak scale,
  and the $C_7^\mathrm{eff}$ function at the hadronic scale.}

\begin{figure*}[t!]
  \epsfxsize 0.32\textwidth
  \figurebox{}{}{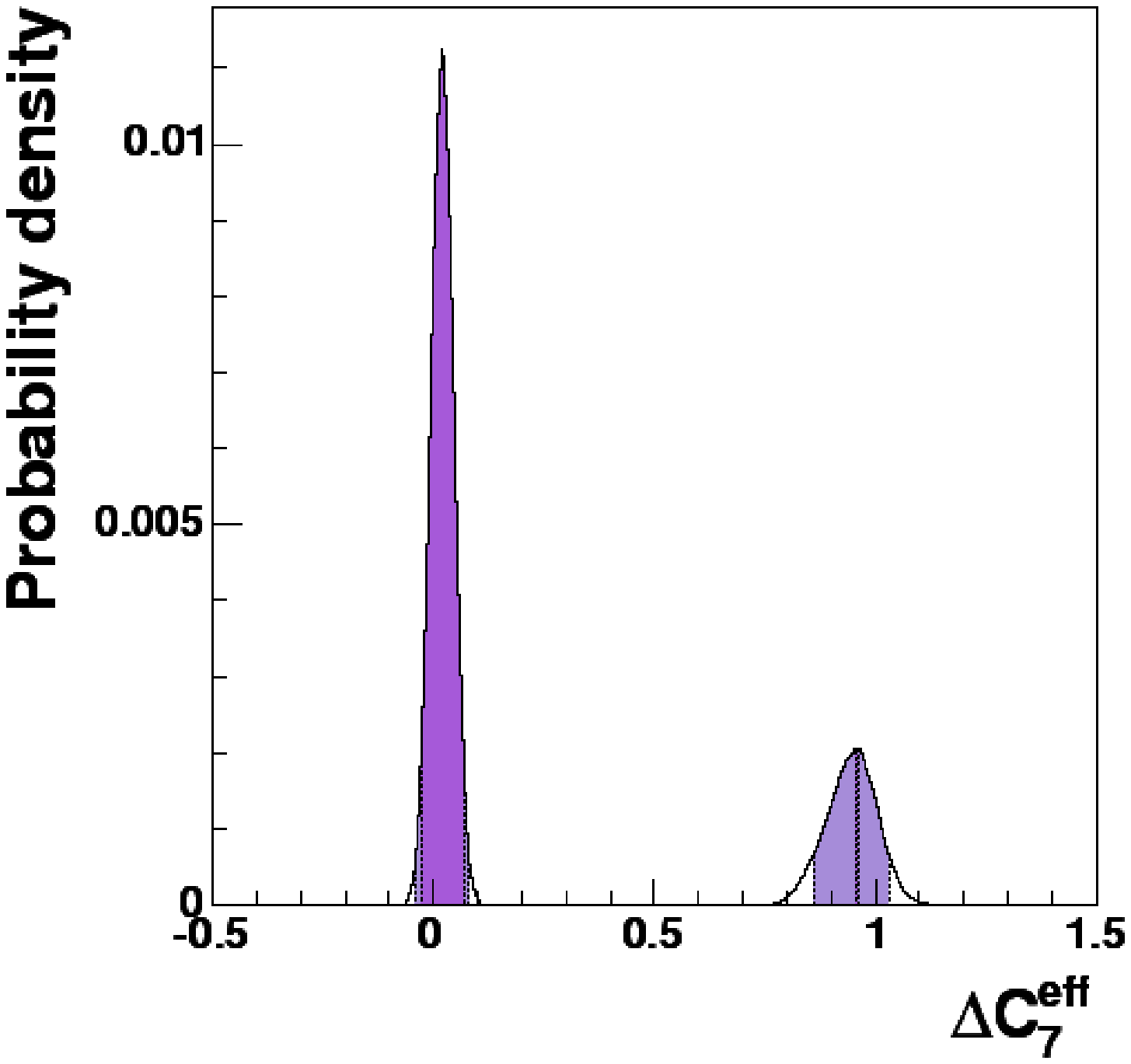}
  \epsfxsize 0.32\textwidth
  \figurebox{}{}{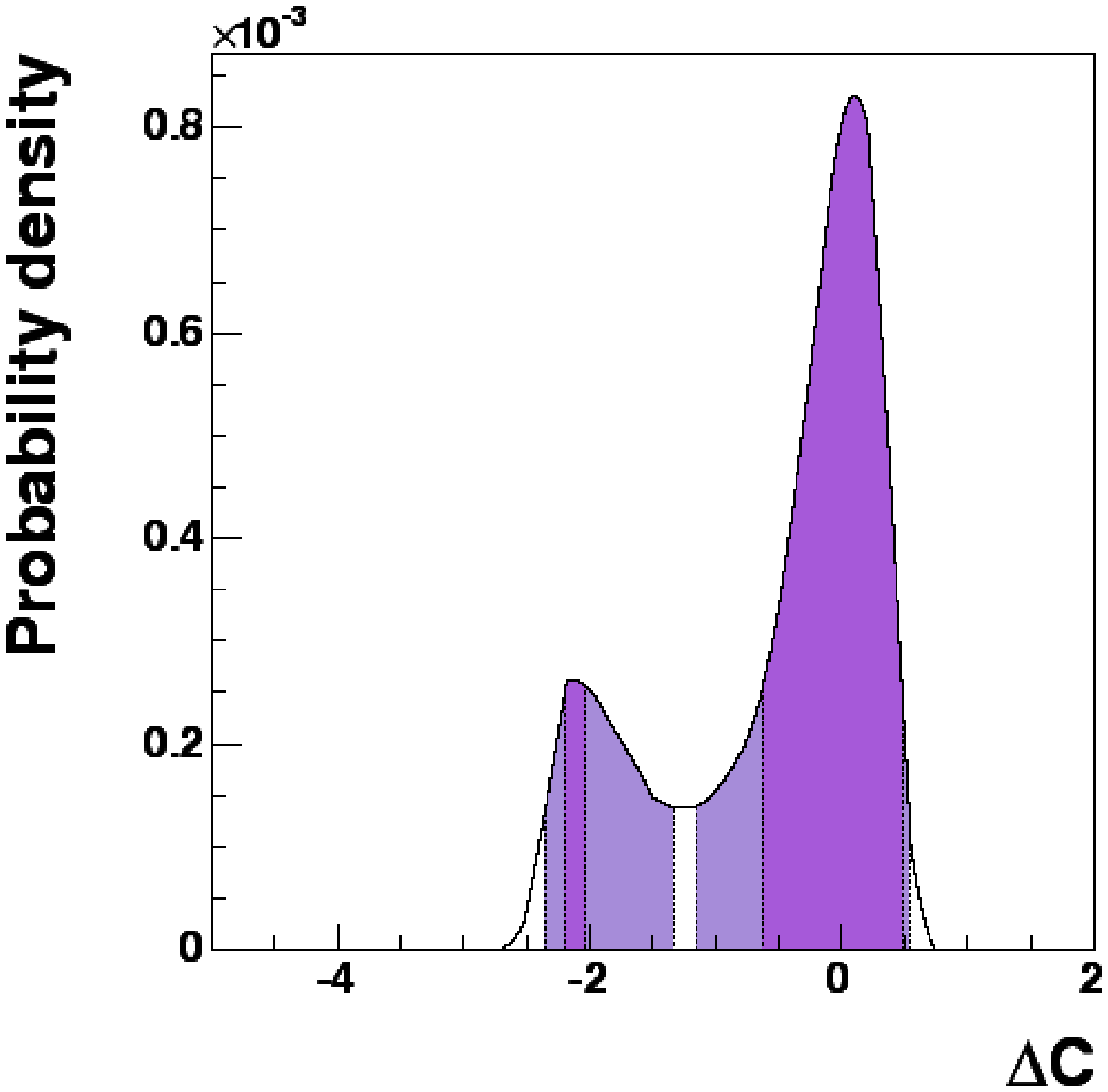}
  \epsfxsize 0.32\textwidth
  \figurebox{}{}{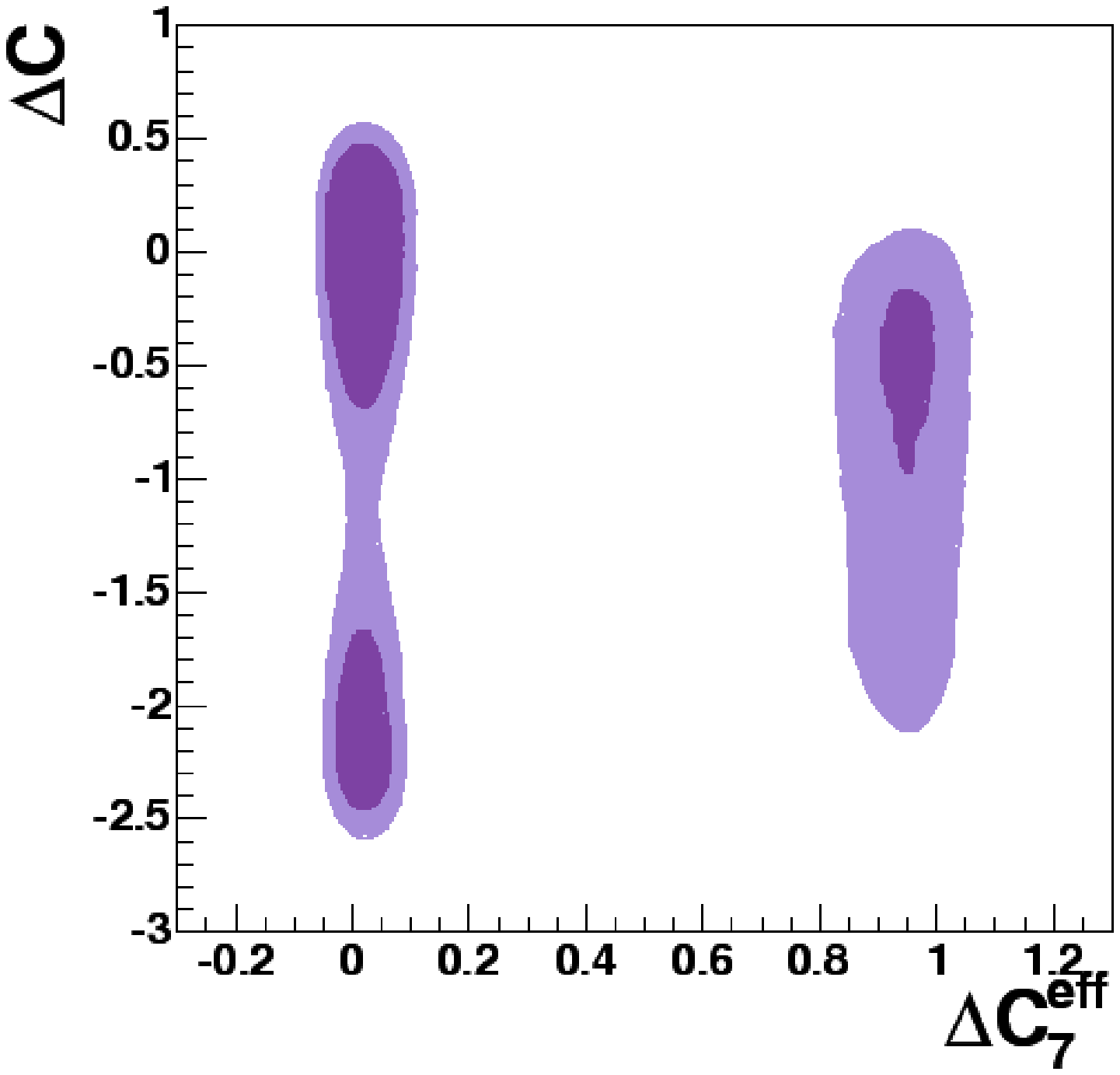}
  \caption{P.d.f.'s for $\Delta C_7^\mathrm{eff}$ (left), $\Delta C$
  (middle) and $\Delta C$ vs.  $\Delta C_7^\mathrm{eff}$ (right).}
  \label{fig:Cs}
\end{figure*}

The analysis goes as follows: using the CKM parameters as determined
by the UUT analysis, one can use BR($B \to X_s \gamma$), BR($B \to X_s
l^+ l^-$) and BR($K^+ \to \pi^+ \nu \bar \nu$) to constrain $\Delta
C$ and $\Delta C_7^\mathrm{eff}$. Then, predictions can be obtained
for all other $K$ and $B$ rare decays. Fig. \ref{fig:Cs} shows the
constraints on the NP contributions. Three possibilities emerge: i)
the SM-like solution with NP corrections close to zero; ii) the
``opposite $C$'' solution with the sign of $C$ flipped by NP and
$C_7^\mathrm{eff}$ close to the SM value; iii) the ``opposite $C_7$''
solution with the sign of $C_7^\mathrm{eff}$ flipped, which however
requires a sizable deviation from the SM also in $C$. 

\begin{figure}[t!]
  \begin{tabular}{cc}
    \epsfxsize 0.22\textwidth
    \figurebox{}{}{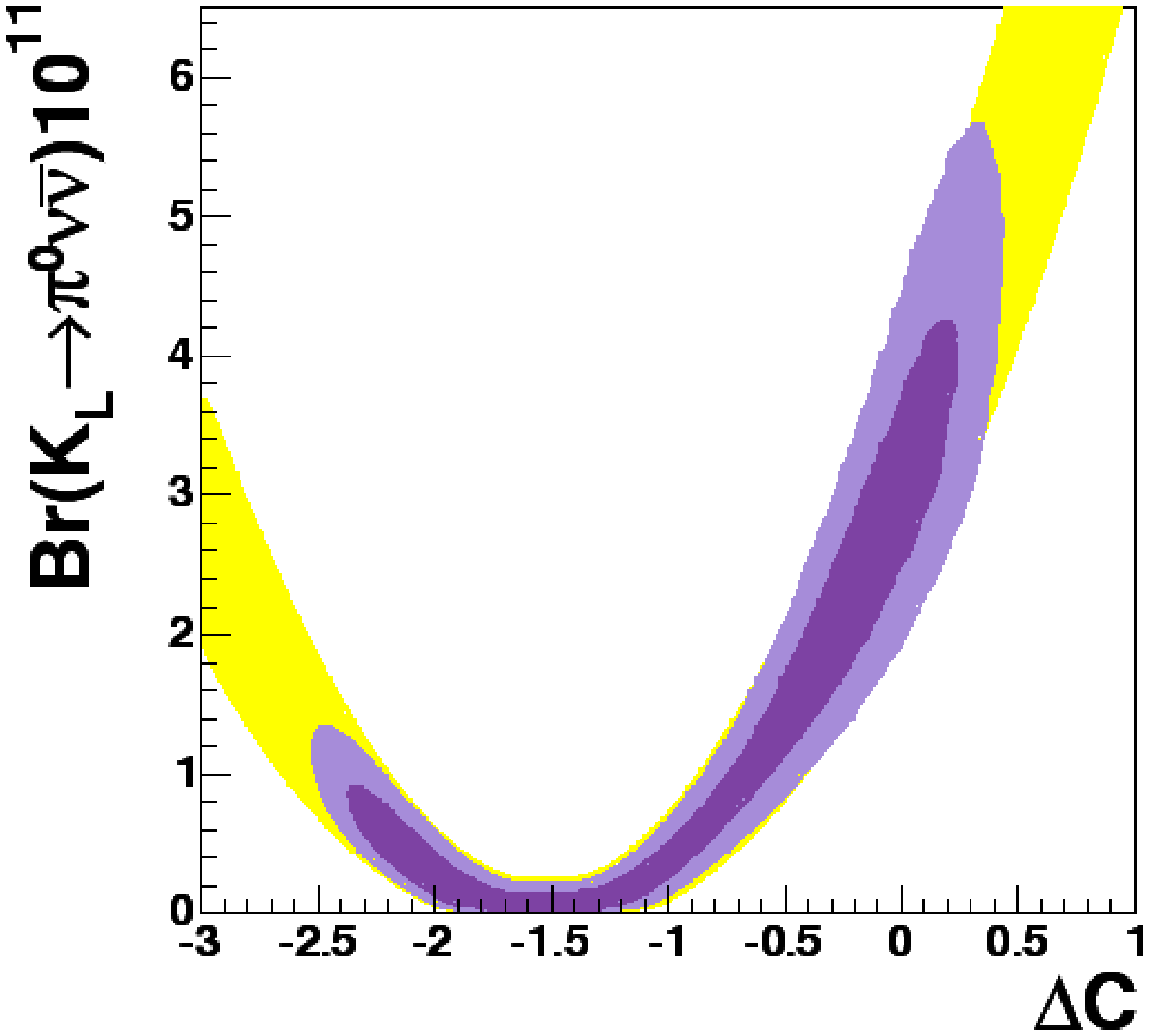} &
    \epsfxsize 0.22\textwidth
    \figurebox{}{}{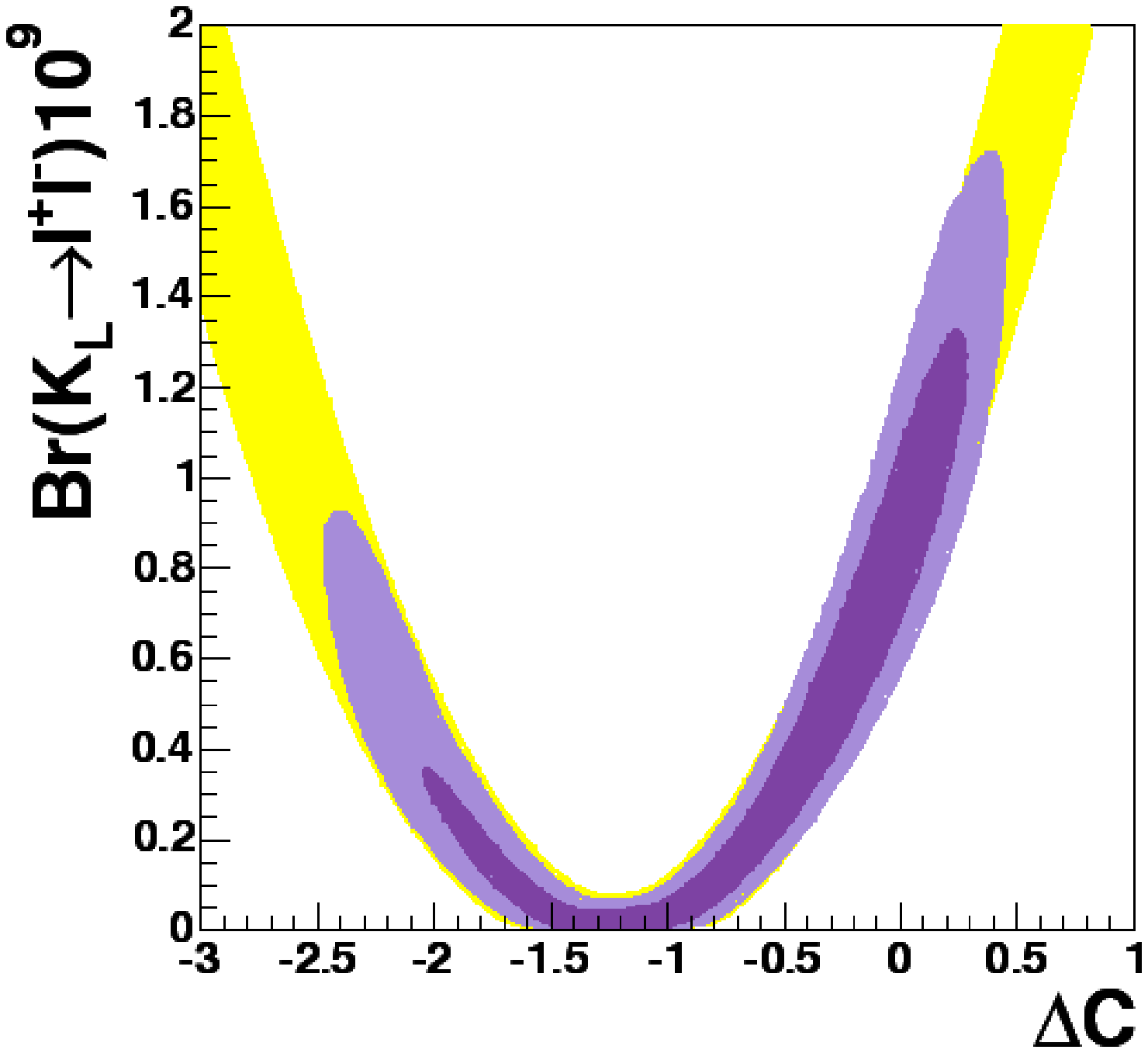} \\
    \epsfxsize 0.22\textwidth
    \figurebox{}{}{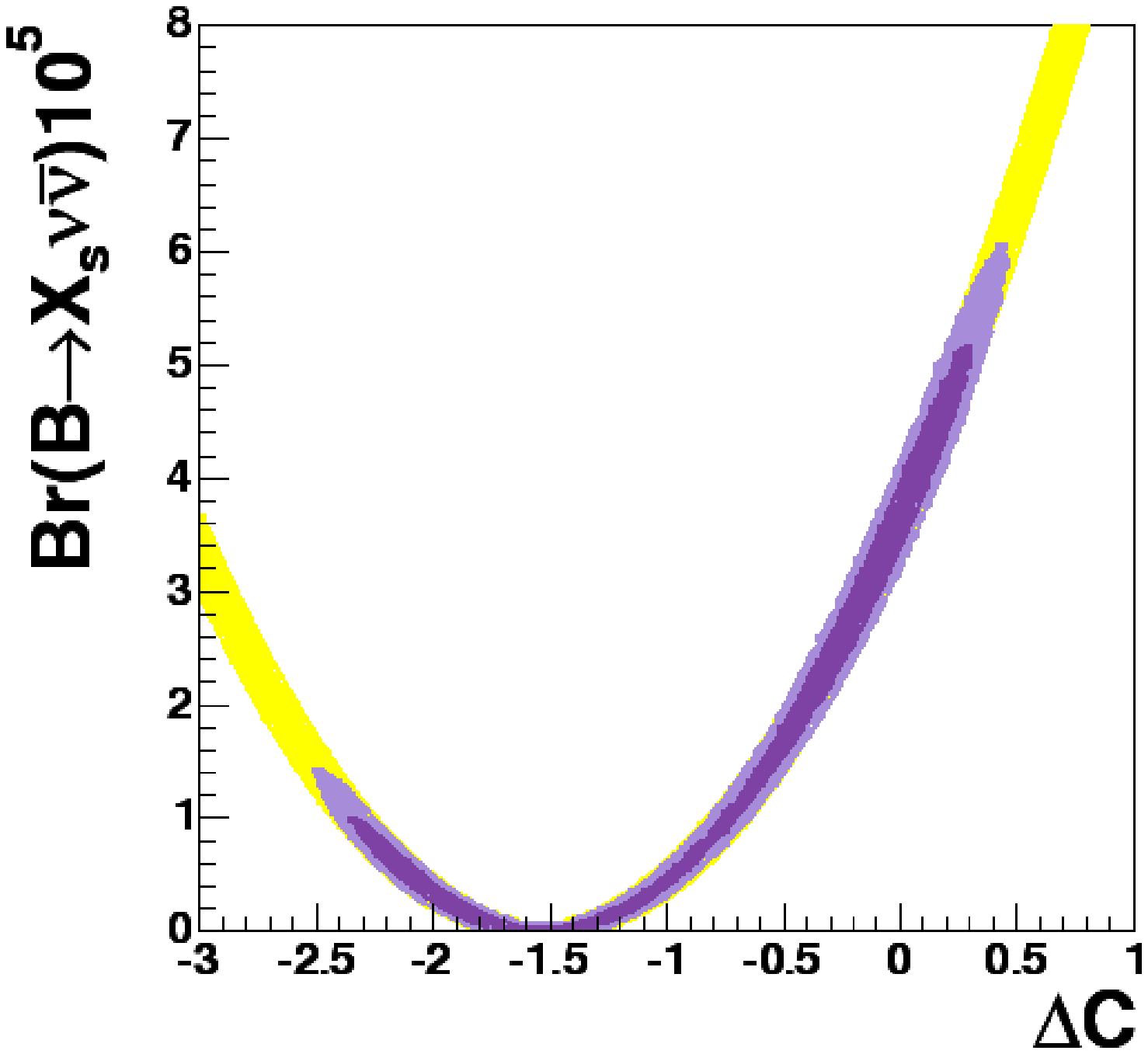} &
    \epsfxsize 0.22\textwidth
    \figurebox{}{}{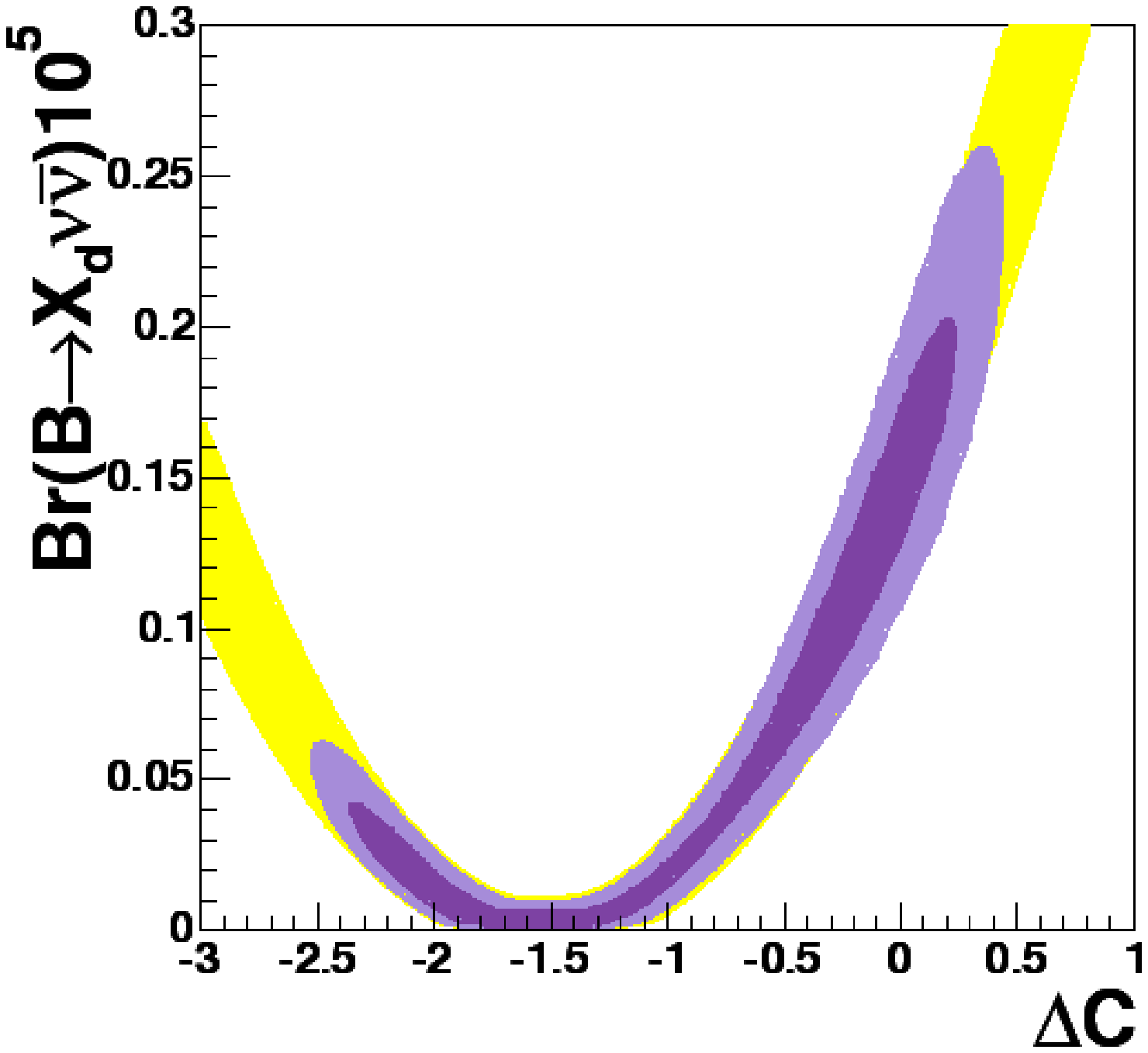} \\
    \epsfxsize 0.22\textwidth
    \figurebox{}{}{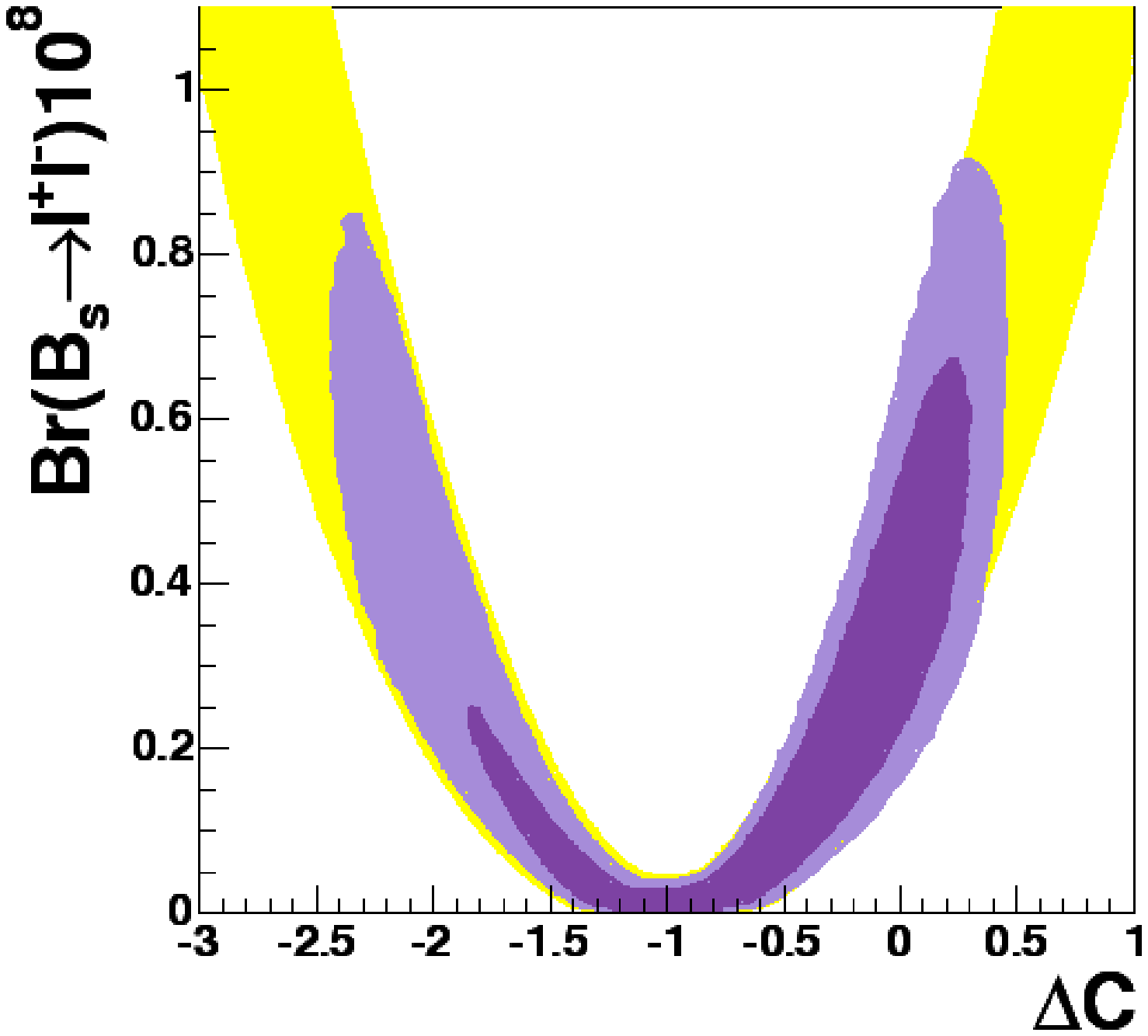} &
    \epsfxsize 0.22\textwidth
    \figurebox{}{}{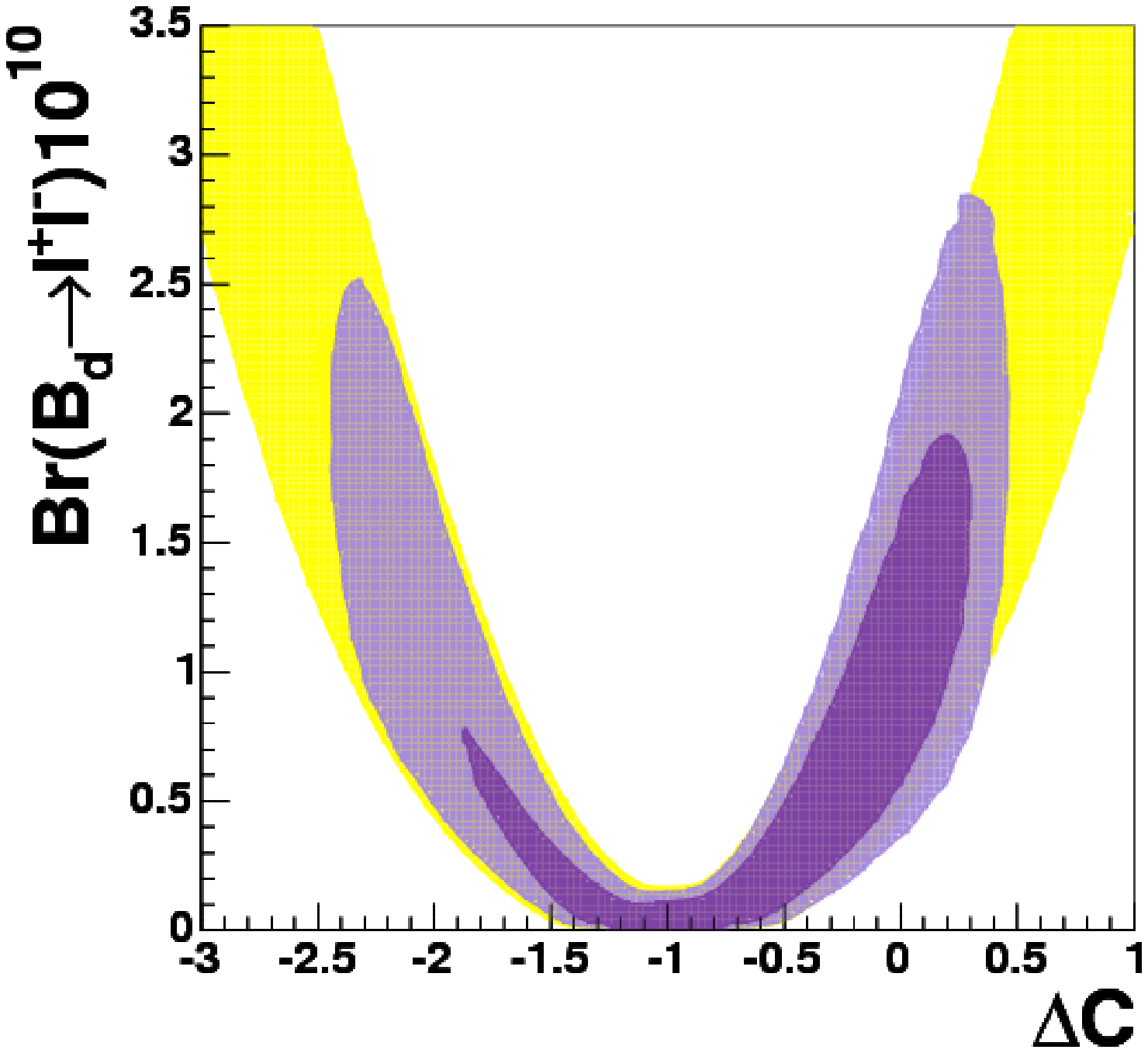}
  \end{tabular}
  \caption{P.d.f.'s for the branching ratios of the rare decays
  $Br(K_L \to \pi^0 \nu \bar \nu)$, $Br(K_L \to \mu \bar \mu)_{\rm
    SD}$, 
  $Br(B\to X_{d,s}\nu\bar\nu)$, and
  $Br(B_{d,s}\to \mu^+\mu^-)$ as a function of $\Delta C$.  Dark
  (light) areas correspond to the $68\%$ ($95\%$) probability region.
  Very light areas correspond to the range obtained without using the
  experimental information.}
  \label{fig:raredec}
\end{figure}

\begin{figure*}[t!]
  \epsfxsize 0.32\textwidth
  \figurebox{}{}{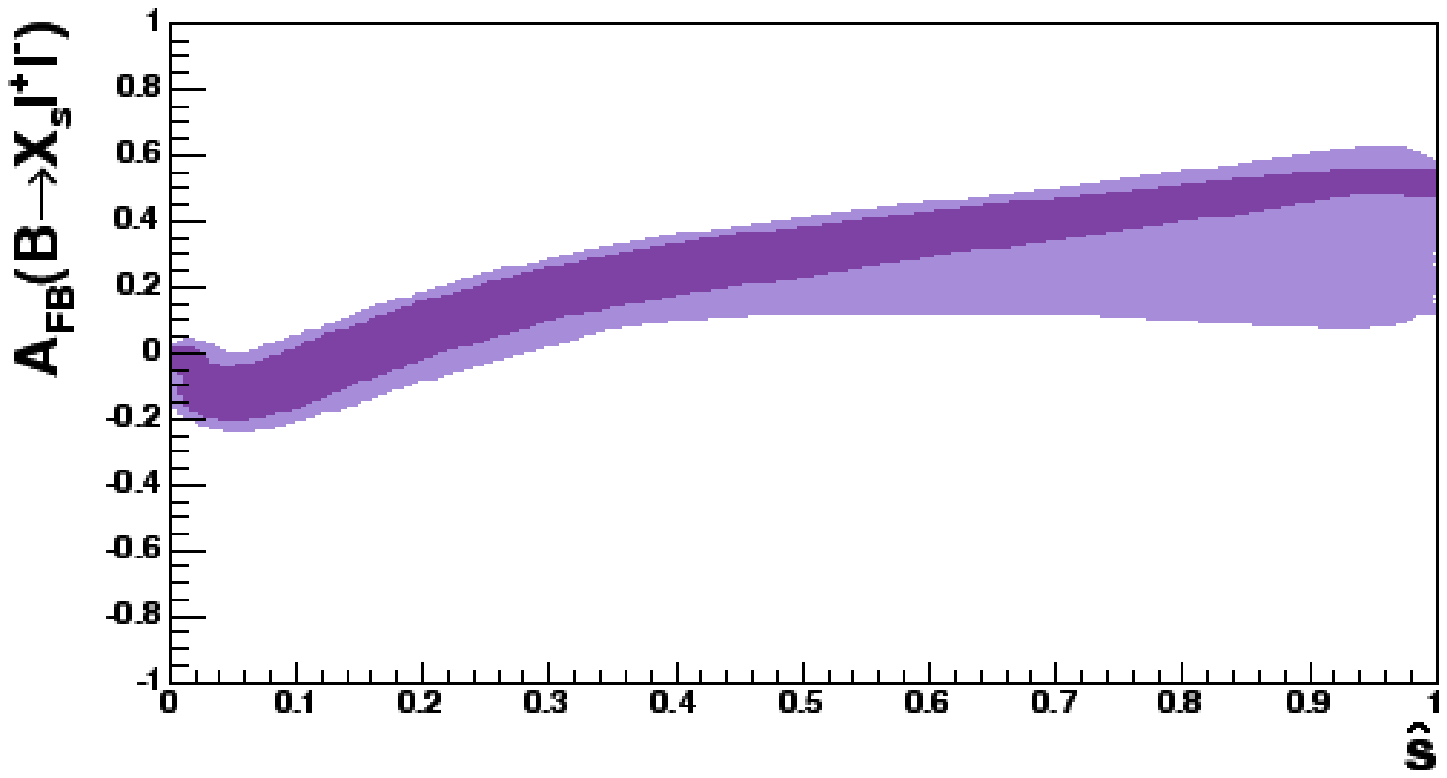}
  \epsfxsize 0.32\textwidth
  \figurebox{}{}{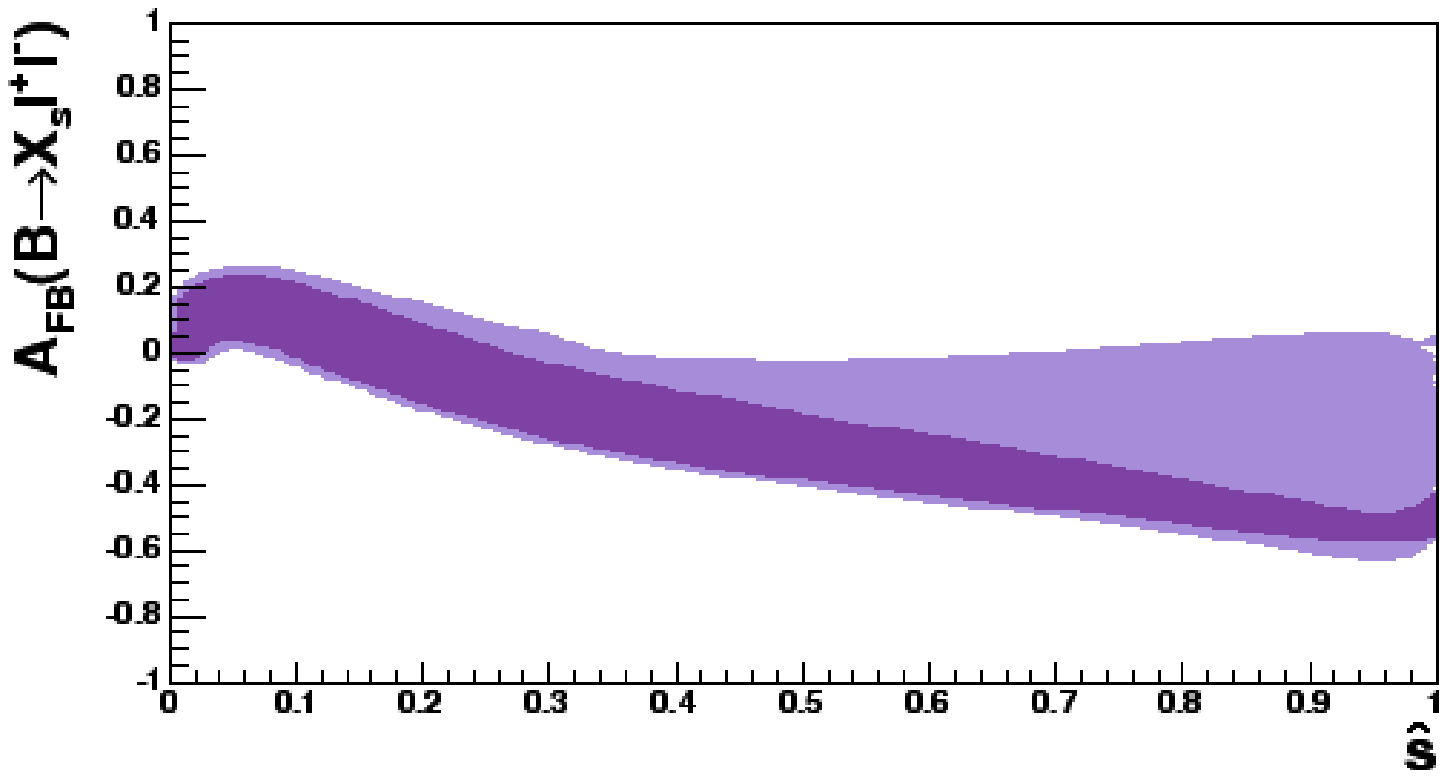}
  \epsfxsize 0.32\textwidth
  \figurebox{}{}{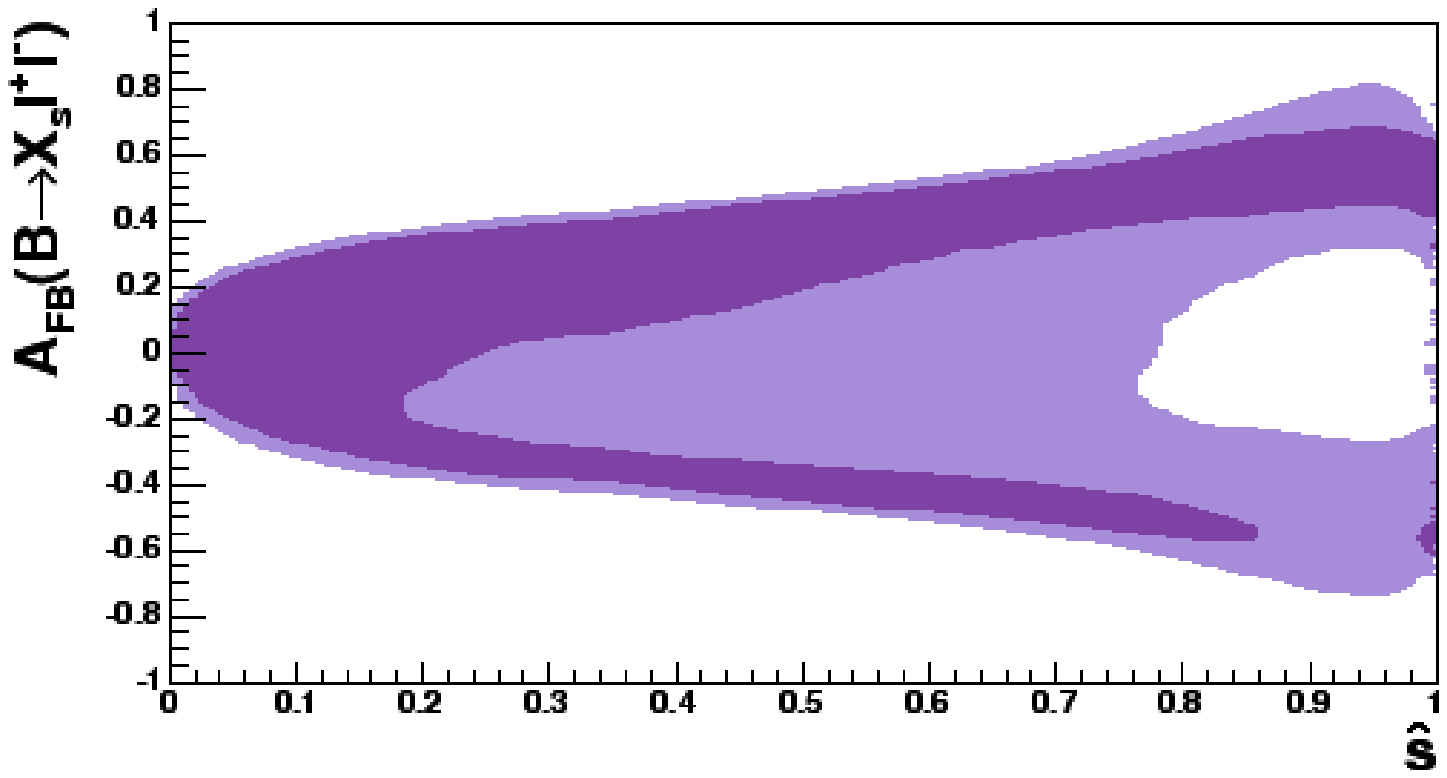}
  \caption{P.d.f. for the normalized forward-backward asymmetry in
    $B \to X_s l^+ l^-$ for $\Delta C_7^\mathrm{eff} \sim 0$
    with $\Delta C> -1$ (left), for $\Delta C_7^\mathrm{eff} \sim 0$
    with $\Delta C < -1$ (middle) and for $\Delta C_7^\mathrm{eff}
    \sim 1$ (right). Dark (light) areas
    correspond to the $68\%$ ($95\%$) probability region.}
  \label{fig:AFB}
\end{figure*}

The corresponding predictions for other rare decays are reported in
Fig.~\ref{fig:raredec}, and the 95\% probability upper bounds are
summarized in Tab.~\ref{tab:MFVBR}, together with the SM predictions
obtained starting from the UUT analysis. It is clear that, given
present constraints, rare decays can be only marginally enhanced with
respect to the SM, while strong suppressions are still possible.
Future improvements in the measurements of BR($B \to X_s \gamma$),
BR($B \to X_s l^+ l^-$) and BR($K^+ \to \pi^+ \nu \bar \nu$) will help
us to reduce the allowed region for NP contributions. Another very
interesting observable is the Forward-Backward asymmetry in $B \to X_s
l^+ l^-$.\cite{AFB} Indeed, the two solutions for $\Delta
C_7^{\mathrm{eff}}$ and the corresponding possible values of $\Delta
C$ give rise to different profiles of the normalized $\bar A_{\rm FB}$
(see eq.~(3.10) of ref.~\cite{BurasMFV}, where more details can be
found).  This can be seen explicitly in Fig.~\ref{fig:AFB}.

\begin{table*}[t!]
  \caption{Upper bounds for rare decays in MFV models at $95 \%$
    probability, the corresponding values in the SM (using inputs from
    the UUT analysis) and the available experimental information.}
\label{tab:MFVBR}
\begin{tabular}{|c|c|c|c|c|}
  \hline
  {Branching Ratios} &  MFV (95\%) &  SM (68\%) &  SM (95\%) & exp\cite{exp}
  \\ \hline
  $Br(K^+ \to \pi^+ \nu \bar \nu)\times 10^{11}$ & $< 11.9$ & $8.3 \pm 1.2$ &  $(6.1,10.9)$
  & $(14.7^{+13.0}_{-8.9})$
  \\ \hline
  $Br(K_L \to \pi^0 \nu \bar \nu)\times 10^{11}$  & $< 4.59$ &  $3.08 \pm 0.56$ &  $(2.03,4.26)$ &
  $ < 5.9 \cdot10^{4}$
  \\ \hline
  $Br(K_L \to \mu \bar \mu)_{\rm SD}\times 10^{9} $ & $< 1.36$ & $0.87 \pm 0.13$ &  $(0.63,1.15)$ & -
  \\ \hline
  $Br(B\to X_s\nu\bar\nu)\times 10^{5}$ & $<5.17$ &  $3.66 \pm 0.21$ &  $(3.25,4.09)$
  &  $<64 $ 
  \\ \hline
  $Br(B\to X_d\nu\bar\nu)\times 10^{6}$ &  $<2.17$ & $1.50 \pm 0.19$ &  $(1.12,1.91)$
  & $<2.2 \cdot 10^2$ 
  \\ \hline
  $Br(B_s\to \mu^+\mu^-)\times 10^{9}$ &  $< 7.42$ & $3.67 \pm 1.01$ &  $(1.91,5.91)$
  & $<1.5\cdot 10^{2}$ 
  \\ \hline
  $Br(B_d\to \mu^+\mu^-)\times 10^{10}$ &  $< 2.20$ & $1.04 \pm 0.34$ &  $(0.47,1.81)$
  & $<3.9 \cdot 10^2$ 
  \\ \hline
\end{tabular}
\end{table*}

\section{New Physics in $\mathbf{b \to s}$ transitions}
\label{sec:b2s}

We concluded sec.~\ref{sec:UTA} pointing out two possible NP scenarios
favoured by the UT analysis: the first one, MFV, was discussed in the
previous section, now we turn to the second one, \textit{i.e.} models
with new sources of flavour and CP violation in $b \to s$ transitions.
Indeed, most NP models fall in this class. Since the SM flavour
$SU(3)$ symmetry is strongly broken by the top (and bottom) Yukawa
couplings, flavour models are not very effective in constraining NP
contributions to $b \to s$ transitions.\cite{piai} The same happens in
models of gauge-Higgs unification or composite Higgs models, due to
the large coupling between the third generation and the EW symmetry
breaking sector.\cite{extra} Last but not least, the large
atmospheric neutrino mixing angle suggests the possibility of large NP
contributions to $b \to s$ processes in SUSY-GUTs.\cite{nuvsq}

This well-motivated scenario is becoming more and more interesting
since $B$ factories are probing NP effects in $b \to s$ penguin
transitions, and the Tevatron and LHCb will probe NP effects in $B_s -
\bar B_s$ mixing in the near future. For the latter process, there is
a solid SM prediction which states that $\Delta m_s > 28\, (30)$
ps$^{-1}$ implies NP at $2 \sigma$ ($3\sigma$). For $b \to s$ penguin
transitions, $B \to X_s \gamma$ and $B \to X_s l^+ l^-$ decays
strongly constrain the FCNC $Z$ and magnetic effective vertices, as
already discussed in the previous section in the simplified case of
MFV. On the other hand, NP contributions to the chromomagnetic $b \to
s$ vertex and to dimension six operators are only mildly constrained
by radiative and semileptonic decays, so that they can contribute
substantially to $b \to s$ hadronic decays, although in any given
model all these NP contributions are in general correlated and thus
more constrained.

As shown in the talk by K.~Abe at this conference, $B$-factories are
now probing NP in $b \to s$ transitions by measuring the coefficient
$\mathcal{S}$ of the $\sin \Delta m_d t$ term in time-dependent CP
asymmetries for $b \to s$ nonleptonic decays. Neglecting the doubly
Cabibbo suppressed $b \to u$ contributions, one should have
$\mathcal{S} = \sin 2 \beta$ for all $b \to s$ channels within the SM,
so that deviations from this equality would signal NP in the decay
amplitude.\cite{npdecay} However, $b \to u$ terms may also cause
deviations $\Delta \mathcal{S}$ from the equality above, so that the
estimate of $\Delta \mathcal{S}$ becomes of crucial importance in
looking for NP. While a detailed analysis of $\Delta \mathcal{S}$ goes
beyond the scope of this talk,\cite{ckmds} the reader should be warned
that $\Delta \mathcal{S}$ might be quite large for channels that are
not pure penguins, and in particular for final states containing
$\eta^\prime$ mesons.~\footnote{Theoretical uncertainties might be
  larger than what expected even in the golden mode $B \to J/\psi
  K_S$, although they can be reduced with the aid of other decay
  modes.\cite{hep-ph/0507290}} In this respect, it is of fundamental
importance to improve the measurement of pure penguin channels, such
as $\phi K_S$, as well as to enlarge the sample of available $b \to s$
and $b \to d$ channels, in order to be able to use flavour symmetries
to constrain $\Delta S$.

The problem of computing $\Delta \mathcal{S}$ in any given NP model is
even tougher: as is well known, in the presence of two contributions
to the amplitude with different weak phases, CP asymmetries depend on
hadronic matrix elements, which at present cannot be computed in a
model-independent way. One has then to resort to models of hadronic
dynamics to estimate $\Delta \mathcal{S}$, with the large theoretical
uncertainties associated to this procedure.

With the above \textit{caveat} in mind, let us now focus on SUSY and
discuss the phenomenological effects of the new sources of flavour and
CP violation in $b \to s$ processes that arise in the squark
sector.\cite{allsusy} In general, in the MSSM squark masses are
neither flavour-universal, nor are they aligned to quark masses, so
that they are not flavour diagonal in the super-CKM basis, in which
quark masses are diagonal and all neutral current (SUSY) vertices are
flavour diagonal. The ratios of off-diagonal squark mass terms to the
average squark mass define four new sources of flavour violation in
the $b \to s$ sector: the mass insertions $(\delta^d_{23})_{AB}$, with
$A,B=L,R$ referring to the helicity of the corresponding quarks. These
$\delta$'s are in general complex, so that they also violate CP. One
can think of them as additional CKM-type mixings arising from the SUSY
sector. Assuming that the dominant SUSY contribution comes from the
strong interaction sector, \textit{i.e.} from gluino exchange, all
FCNC processes can be computed in terms of the SM parameters plus the
four $\delta$'s plus the relevant SUSY masses: the gluino mass
$m_{\tilde g}$, the average squark mass $m_{\tilde q}$ and, in
general, $\tan \beta$ and the $\mu$ parameter.\footnote{the last two
  parameters are irrelevant as long as $\tan \beta$ is of
  $\mathcal{O}(1)$.} Barring accidental cancellations, one can
consider one single $\delta$ parameter, fix the SUSY masses and study
the phenomenology. The constraints on $\delta$'s come at present from
BR's and CP asymmetries in $B \to X_s \gamma$, $B \to X_s l^+ l^-$ and
from the lower bound on $\Delta m_s$. Since gluino exchange does not
generate a sizable $\Delta C$ in the notation of the previous section,
the combined constraints from radiative and semileptonic decays are
particularly stringent.

\begin{figure}[h!]
  \begin{tabular}{cc}
    \epsfxsize 0.24\textwidth
    \figurebox{}{}{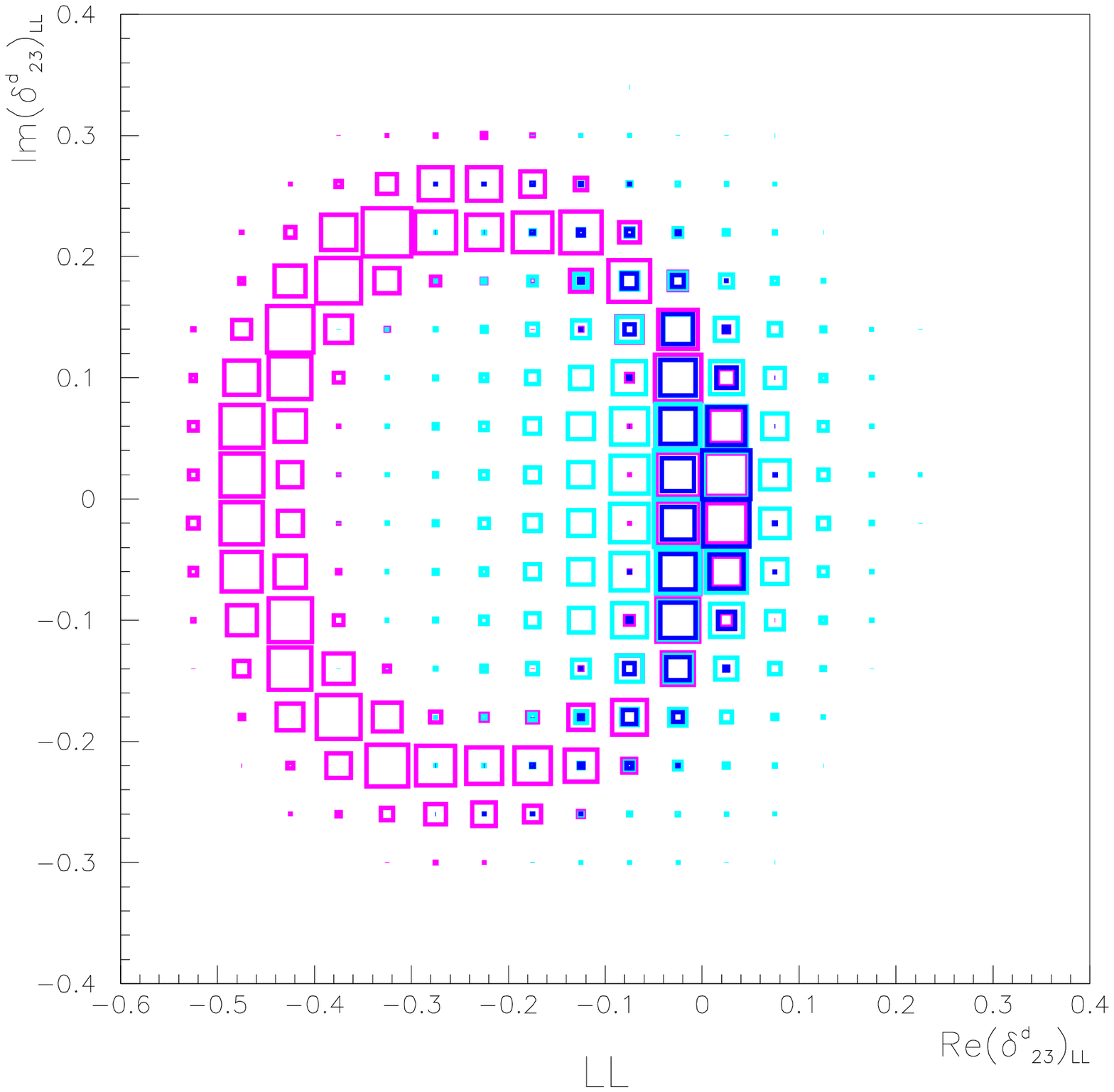} &
    \epsfxsize 0.24\textwidth
    \figurebox{}{}{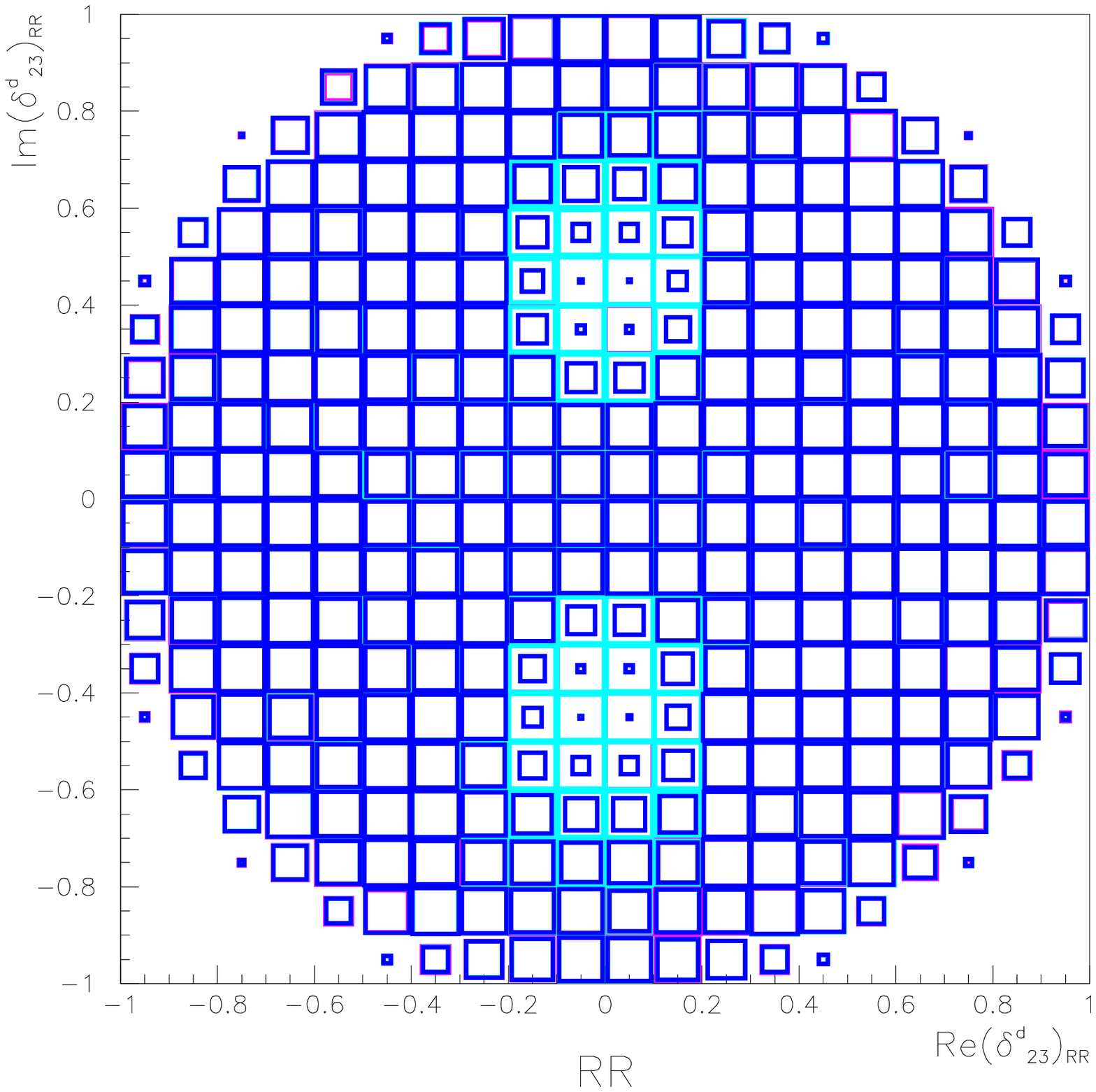} \\
    \epsfxsize 0.24\textwidth
    \figurebox{}{}{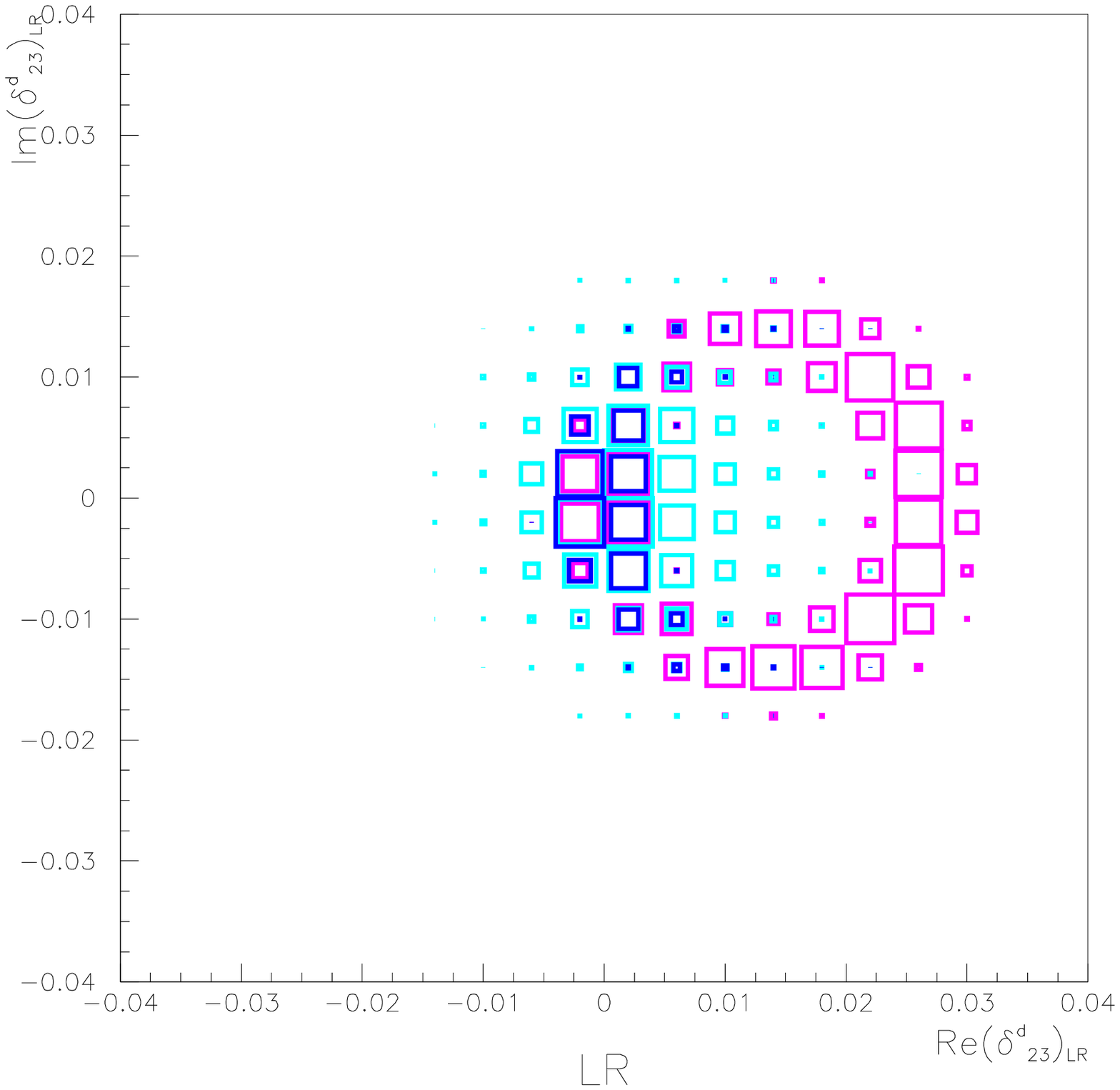} &
    \epsfxsize 0.24\textwidth
    \figurebox{}{}{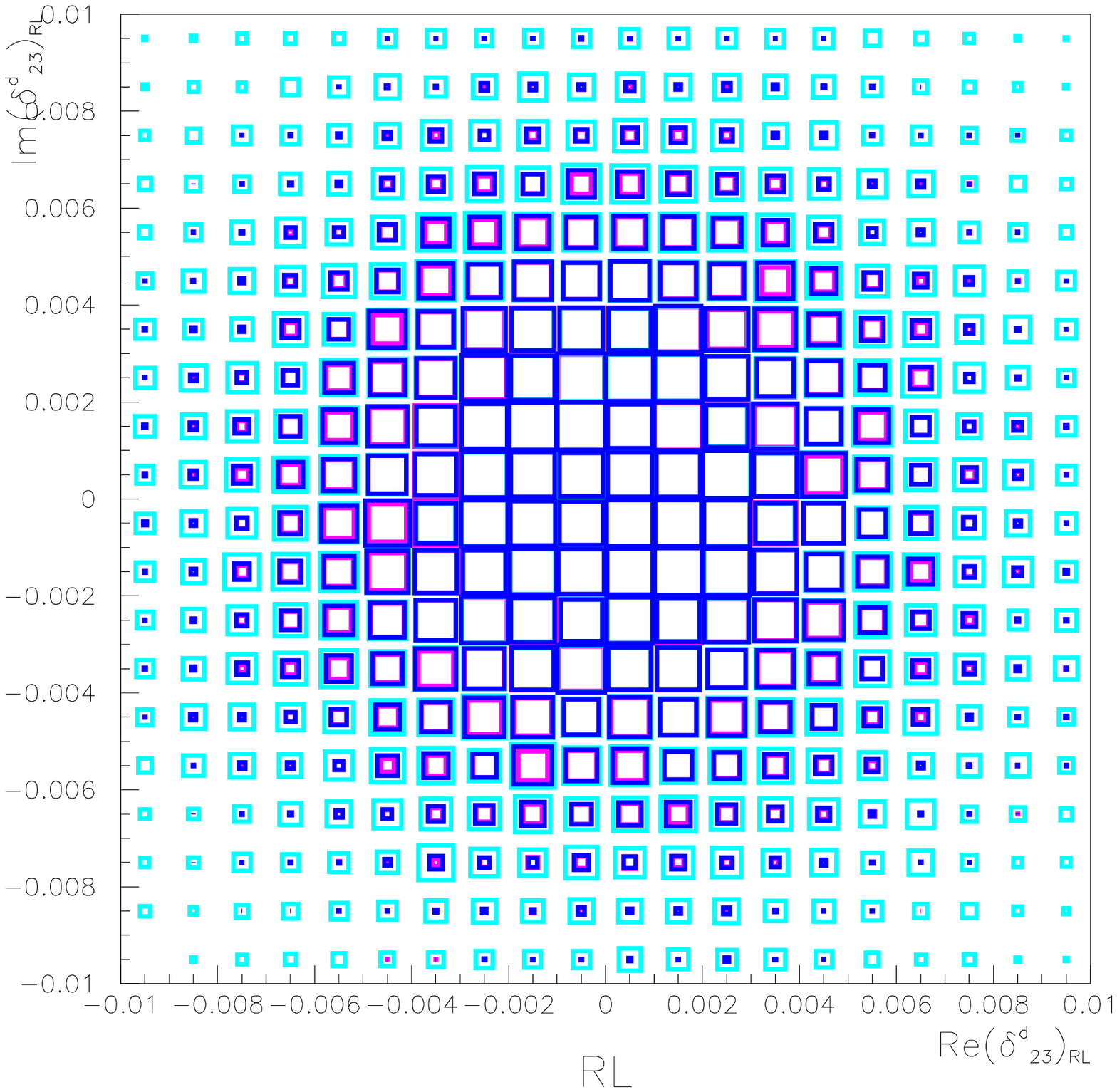} 
  \end{tabular}
  \caption{P.d.f.'s in the $\mathrm{Re} (\delta^d_{23})_{AB} -
    \mathrm{Im} (\delta^d_{23})_{AB}$ plane for $A,B=L,R$, as
    determined by $B \to X_s \gamma$ (violet), $B \to X_s l^+ l^-$
    (light blue) and all constraints (dark blue).}
  \label{fig:susyconstr}
\end{figure}

Fixing as an example $m_{\tilde g}=m_{\tilde q}= - \mu =$ 350 GeV and
$\tan \beta = 10$, one obtains the constraints on $\delta$'s reported
in Fig.~\ref{fig:susyconstr}. Several comments are in order at this
point: i) only $(\delta^d_{23})_{\mathrm{LL},\mathrm{LR}}$ generate
amplitudes that interfere with the SM one in rare decays. Therefore,
the constraints from rare decays for
$(\delta^d_{23})_{\mathrm{RL},\mathrm{RR}}$ are symmetric around zero,
while the interference with the SM produces the circular shape of the
$B \to X_s \gamma$ constraint on
$(\delta^d_{23})_{\mathrm{LL},\mathrm{LR}}$. ii) We recall that
$\mathrm{LR}$ and $\mathrm{RL}$ mass insertions generate much larger
contributions to the (chromo)magnetic operators, since the necessary
chirality flip can be performed on the gluino line ($\propto m_{\tilde
  g}$) rather than on the quark line ($\propto m_{\tilde b}$).
Therefore, the $B \to X_s \gamma$ constraint is much more effective on
these insertions. iii) The $\mu \tan \beta$ flavour-conserving
$\mathrm{LR}$ squark mass term generates, together with a flavour
changing $\mathrm{LL}$ mass insertion, an effective
$(\delta^d_{23})_{\mathrm{LR}}^\mathrm{eff}$ that contributes to $B
\to X_s \gamma$. Having chosen a negative $\mu$, we have
$(\delta^d_{23})_{\mathrm{LR}}^\mathrm{eff} \propto -
(\delta^d_{23})_{\mathrm{LL}}$ and therefore the circle determined by
$B \to X_s \gamma$ in the $\mathrm{LL}$ and $\mathrm{LR}$ cases lies
on opposite sides of the origin (see Fig.~\ref{fig:susyconstr}). iv)
For $\mathrm{LL}$ and $\mathrm{LR}$ cases, $B \to X_s \gamma$ and $B
\to X_s l^+ l^-$ produce bounds with different shapes on the Re
$\delta$ -- Im $\delta$ plane (violet and light blue regions in
Fig.~\ref{fig:susyconstr}), so that applying them simultaneously only
a much smaller region around the origin survives (dark blue regions in
Fig.~\ref{fig:susyconstr}). This shows the key role played by rare
decays in constraining new sources of flavour and CP violation in the
squark sector. v) for the $\mathrm{RR}$ case, the constraints from
rare decays are very weak, so that almost all $\delta$'s with
$\vert(\delta^d_{23})_{\mathrm{RR}}\vert < 1$ are allowed, except for
two small forbidden regions where $\Delta m_s$ goes below the
experimental lower bound.

\begin{figure}[h!]
  \begin{tabular}{cc}
    \epsfxsize 0.24\textwidth
    \figurebox{}{}{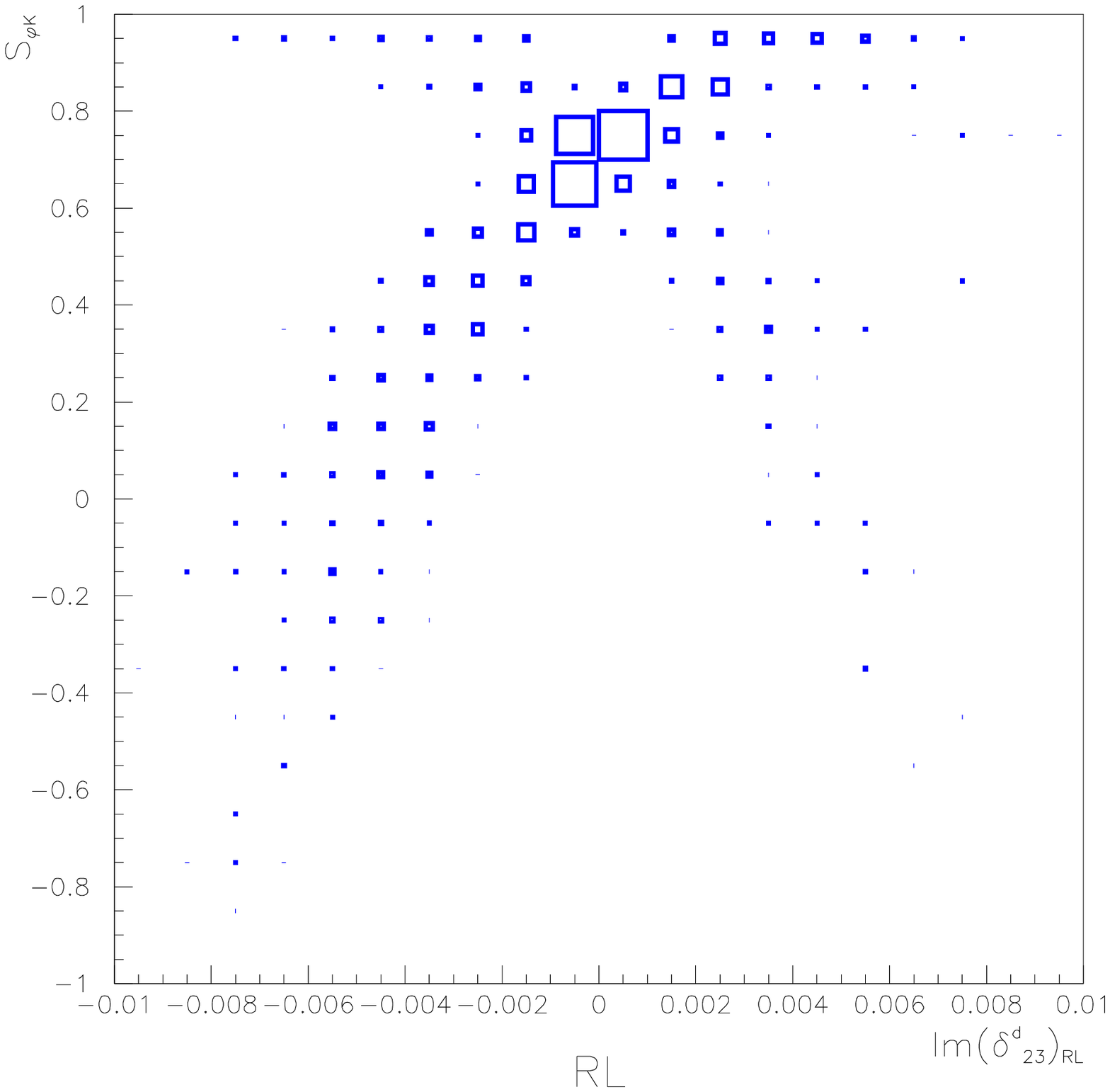} &
    \epsfxsize 0.24\textwidth
    \figurebox{}{}{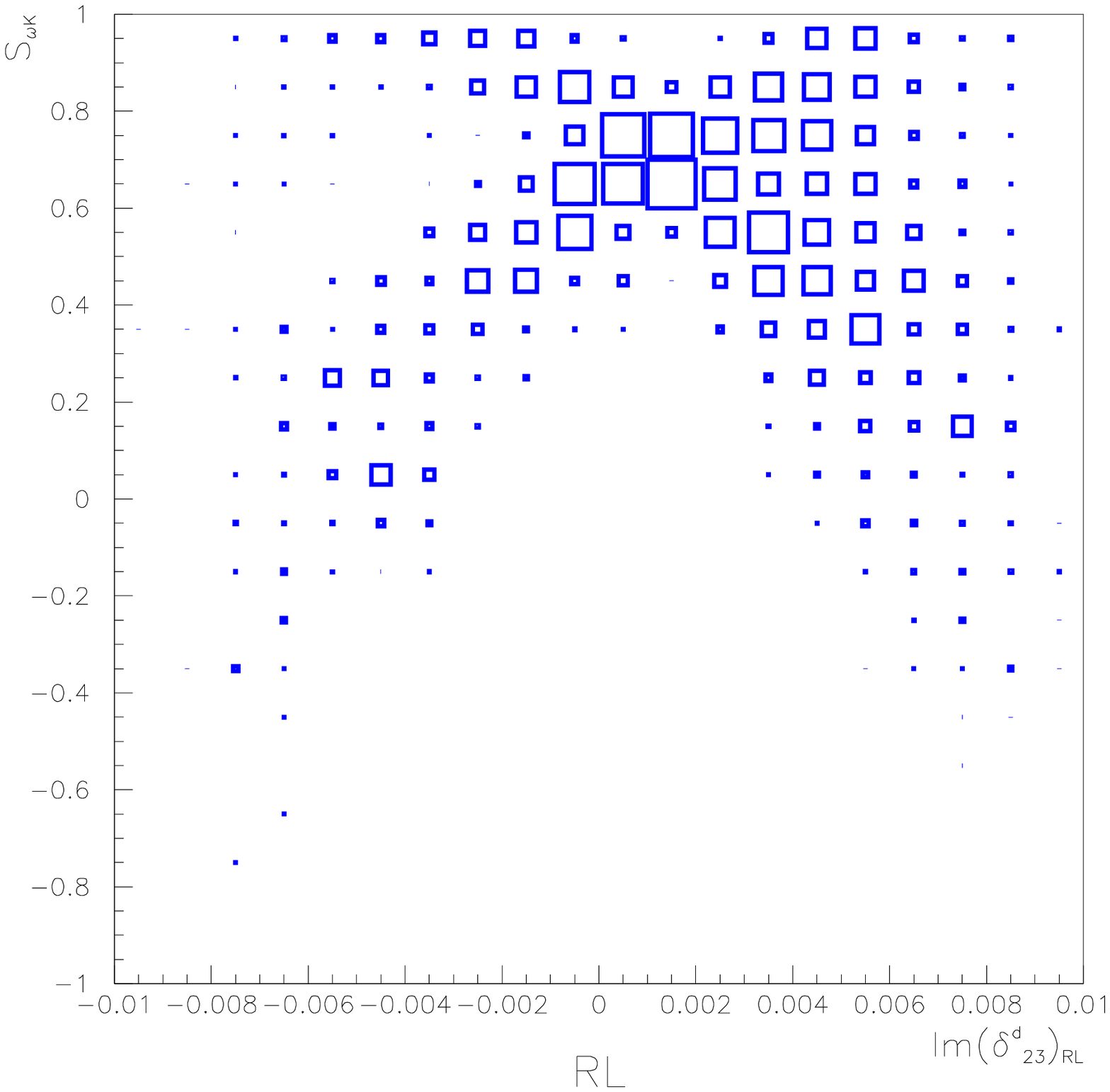} \\
    \epsfxsize 0.24\textwidth
    \figurebox{}{}{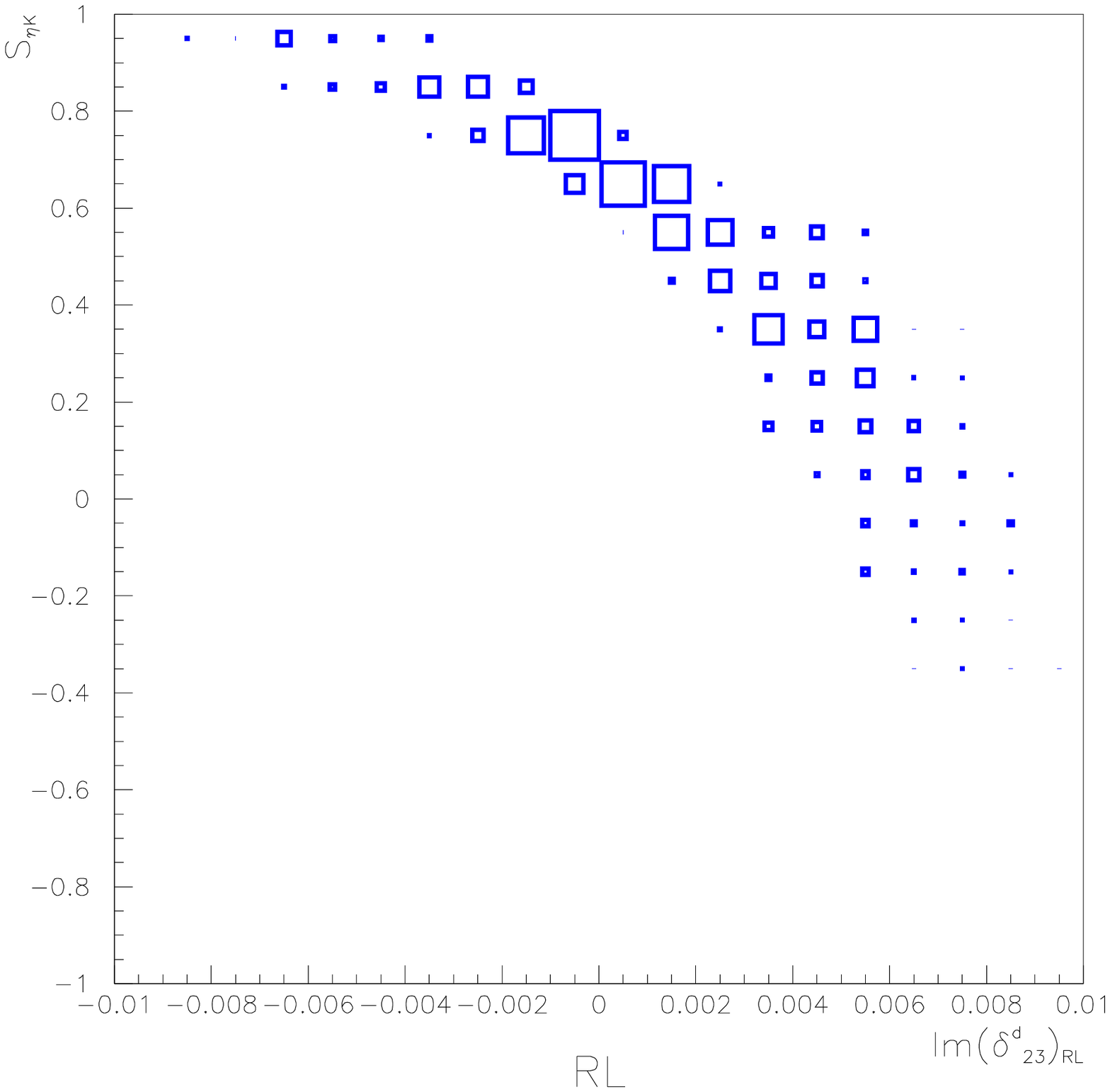} &
    \epsfxsize 0.24\textwidth
    \figurebox{}{}{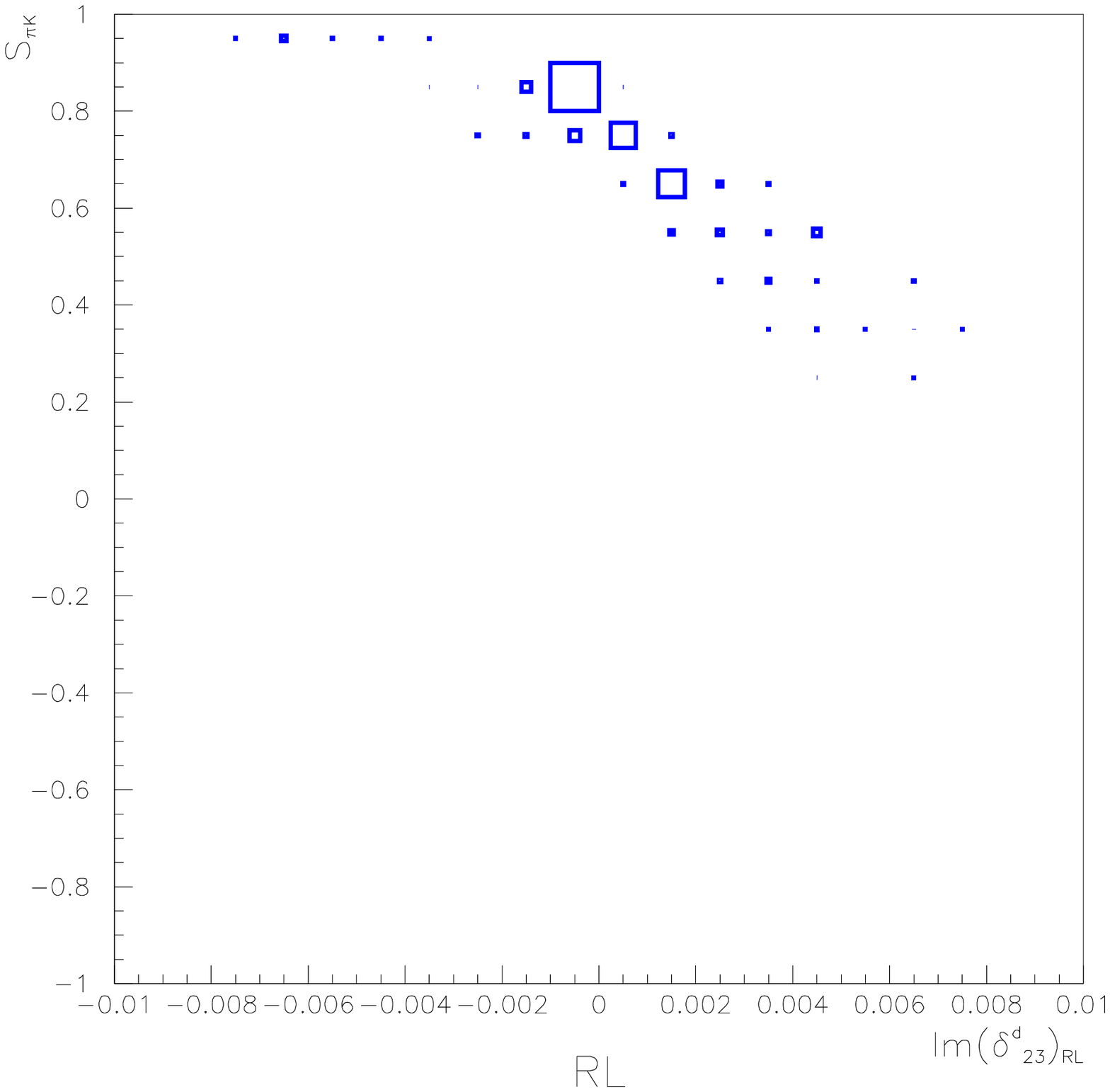} 
  \end{tabular}
  \caption{From top to bottom and from left to right, p.d.f.'s for
    $\mathcal{S}$ for $B$ decays to $\phi K_S$, $\omega K_S$,
    $\eta^\prime K_S$ and $\pi K_S$ as a function of Im
    $(\delta^d_{23})_{\mathrm{RL}}$.}
  \label{fig:susyS}
\end{figure}

Having determined the p.d.f's for the four $\delta$'s, we now turn to
the evaluation of $\mathcal{S}$ as defined at the beginning of this
section. We use the approach defined in ref.~\cite{CFMS} to evaluate
the relevant hadronic matrix elements, warning the reader about the
large uncontrolled theoretical uncertainties that affect this
evaluation. Let us focus for concreteness on the effects of
$(\delta^d_{23})_{\mathrm{RL}}$. Imposing that the $BR$'s are
correctly reproduced, we obtain the estimates of $\mathcal{S}$ for the
$\phi K_s$, $\eta^\prime K_s$, $\omega K_s$ and $\pi^0 K_s$ final
states reported in Fig.~\ref{fig:susyS}. One can see that
$(\delta^d_{23})_{\mathrm{RL}}$ insertions can produce sizable
deviations from the SM expectations for $\mathcal{S}$ in the
$\eta^\prime K_s$ and $\omega K_s$ channels. Similar results hold for
the other $\delta$'s.

\begin{figure*}[t!]
    \epsfxsize 0.45\textwidth
    \figurebox{}{}{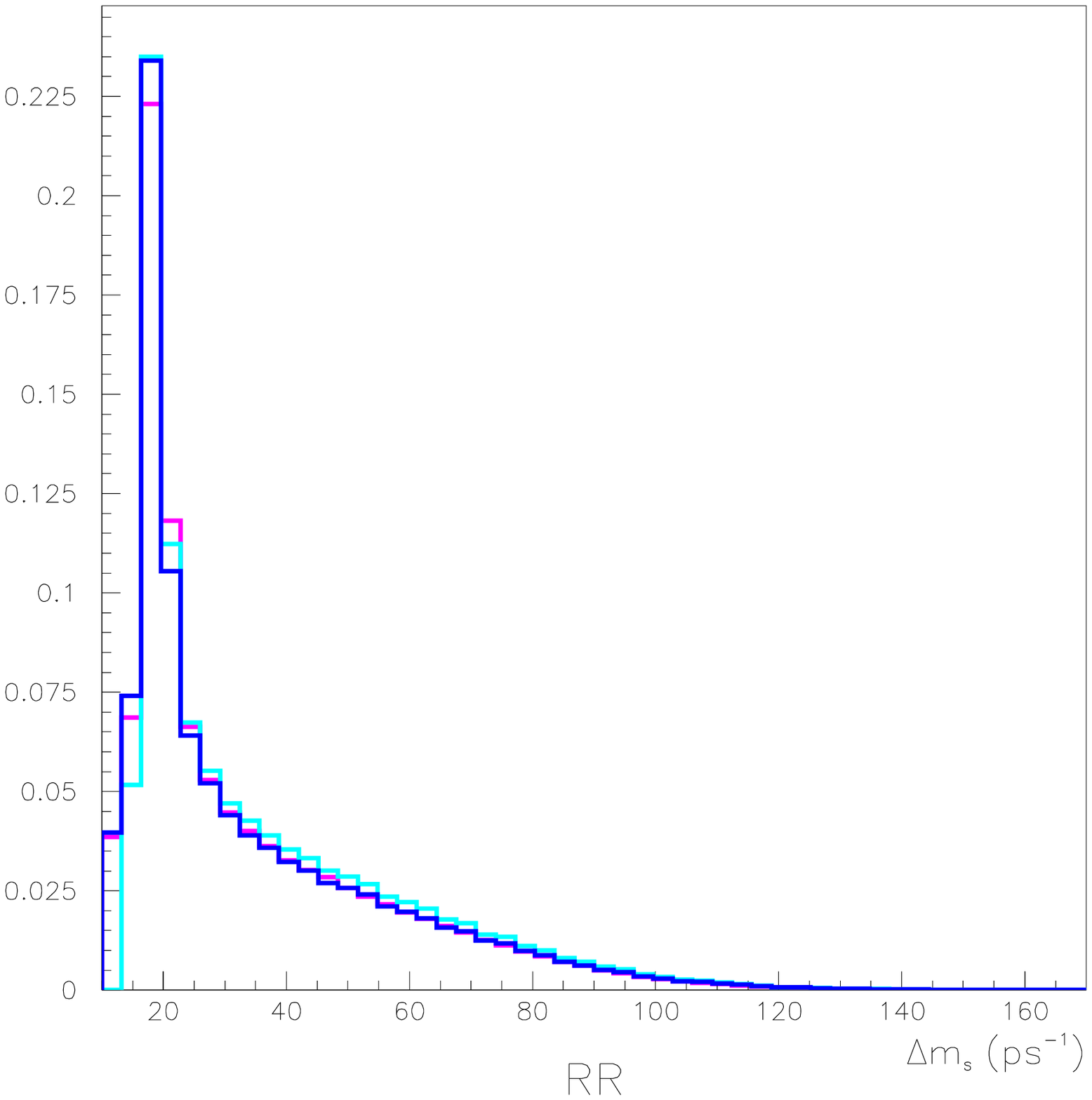}
    \epsfxsize 0.45\textwidth
    \figurebox{}{}{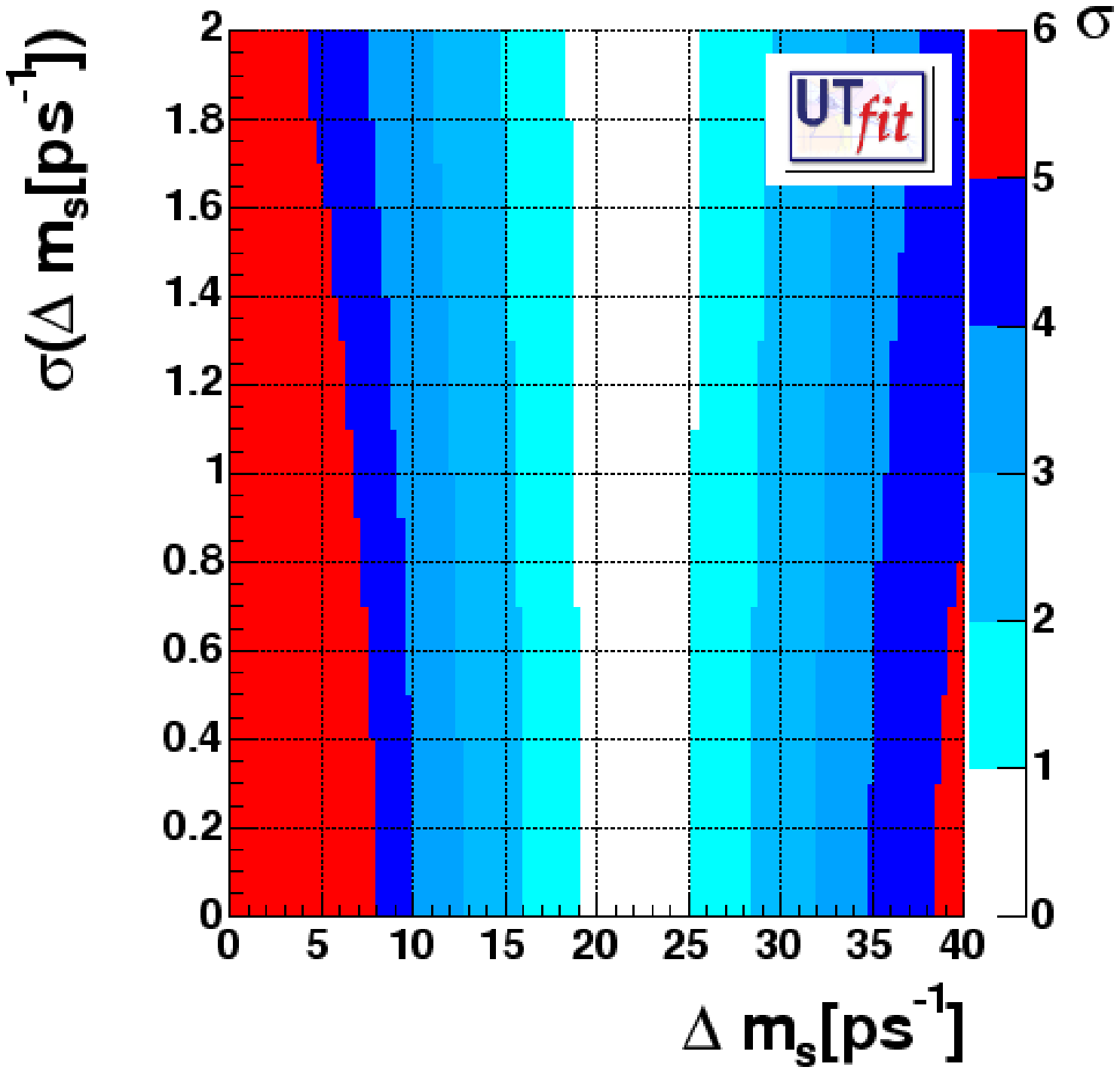}
  \caption{Left: p.d.f. for $\Delta m_S$ obtained considering
    $(\delta^d_{23})_{\mathrm{RR}}$ and the SUSY parameters given in the text,
    as determined by $B \to X_s \gamma$ (violet), $B \to X_s l^+
    l^-$ (light blue) and all constraints (dark blue). Right:
    compatibility plot for $\Delta m_s$ in the SM.}
  \label{fig:susydms}
\end{figure*}

Another place where $\delta^d_{23}$ mass insertions can produce large
deviations from the SM is $\Delta m_s$. In this case, hadronic
uncertainties are under control, thanks to the Lattice QCD computation
of the relevant matrix elements,\cite{hep-lat/0110091} and the whole
computation is at the same level of accuracy of the SM
one.\cite{SUSYNLO} Considering for example the contribution of
$(\delta^d_{23})_{\mathrm{RR}}$ mass insertions, starting from the
constraints in Fig.~\ref{fig:susyconstr}, one obtains the p.d.f. for
$\Delta m_s$ reported in Fig.~\ref{fig:susydms}, where for comparison
we also report the compatibility plot within the SM.\cite{utfit2005}
Much larger values are possible in the SUSY case, generally
accompanied by large values of the CP asymmetry in $B_s \to J/\psi
\phi$: both would be a clear signal of NP to be revealed at hadron
colliders.

\begin{figure*}[t!]
  \begin{tabular}{cc}
    \epsfxsize 0.45\textwidth
    \figurebox{}{}{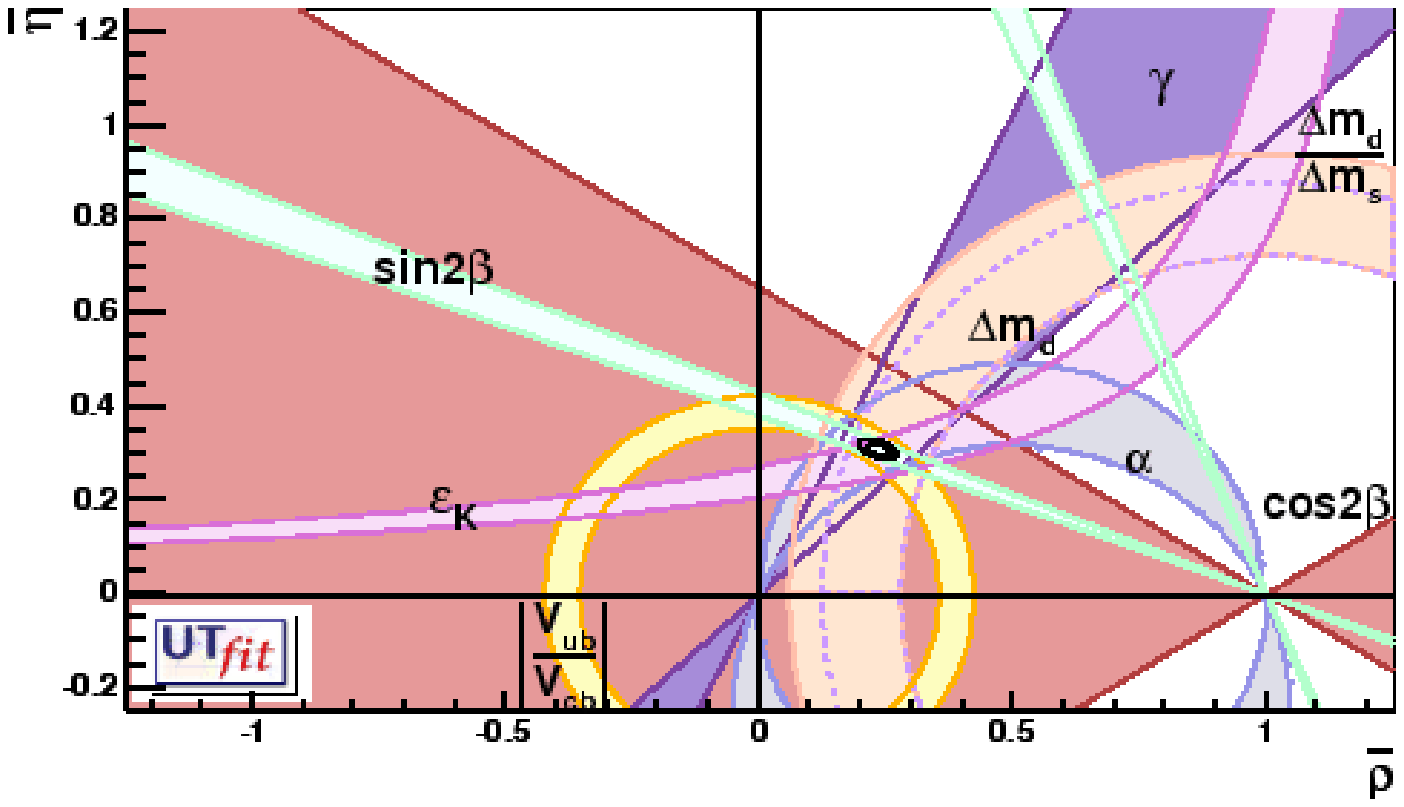} &
    \epsfxsize 0.35\textwidth
    \figurebox{}{}{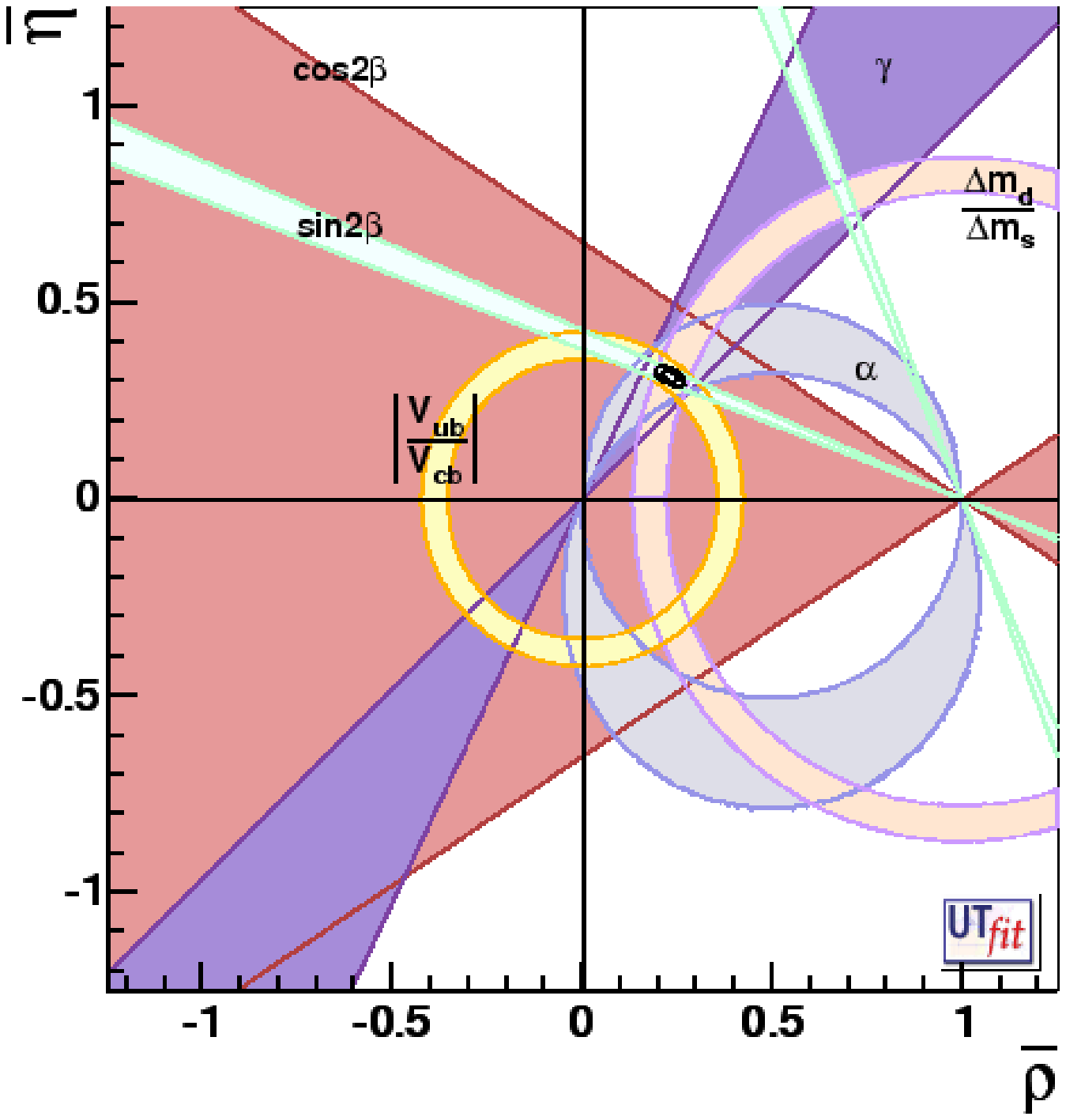} \\
    \epsfxsize 0.4\textwidth
    \figurebox{}{}{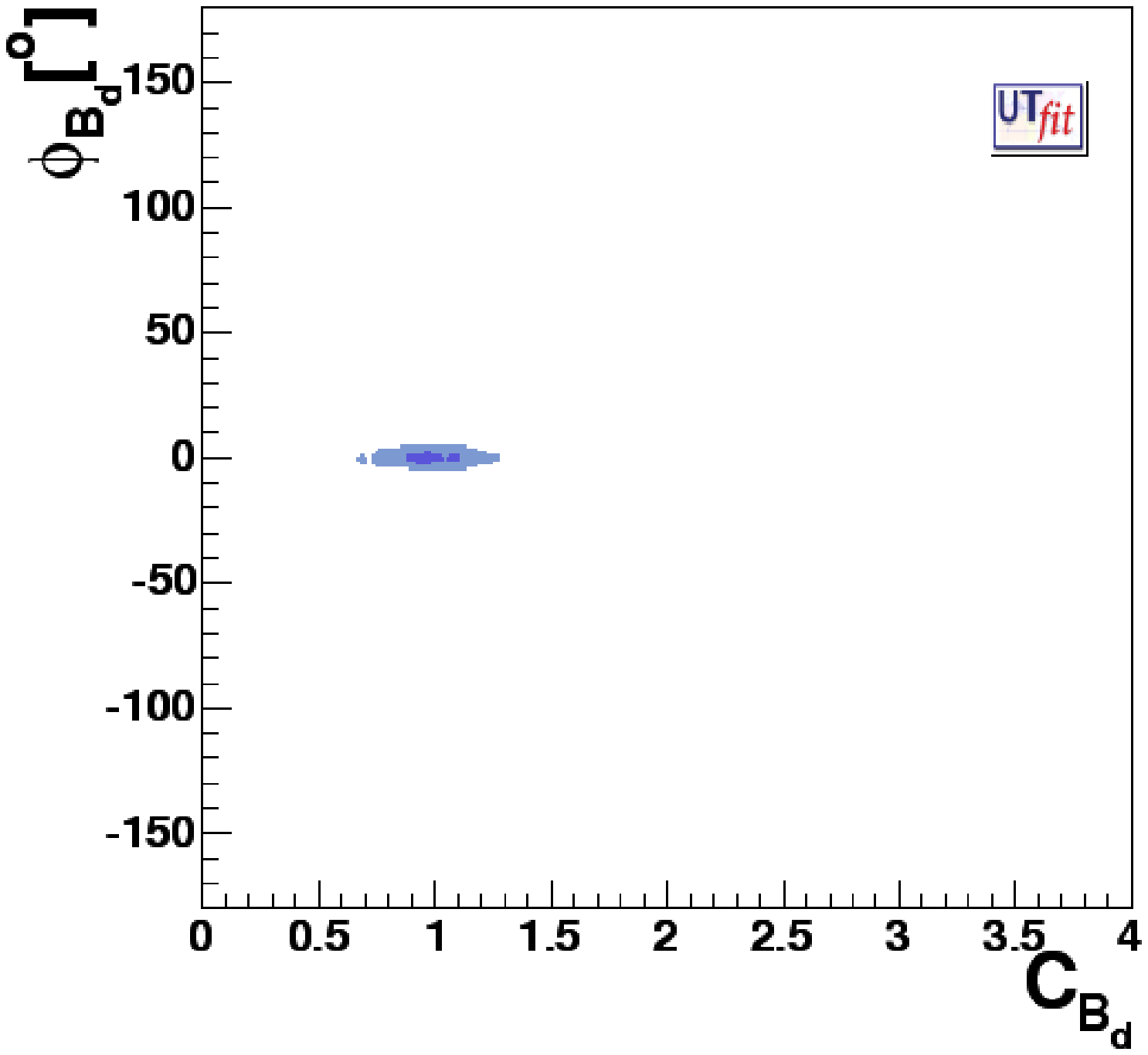} &
    \epsfxsize 0.4\textwidth
    \figurebox{}{}{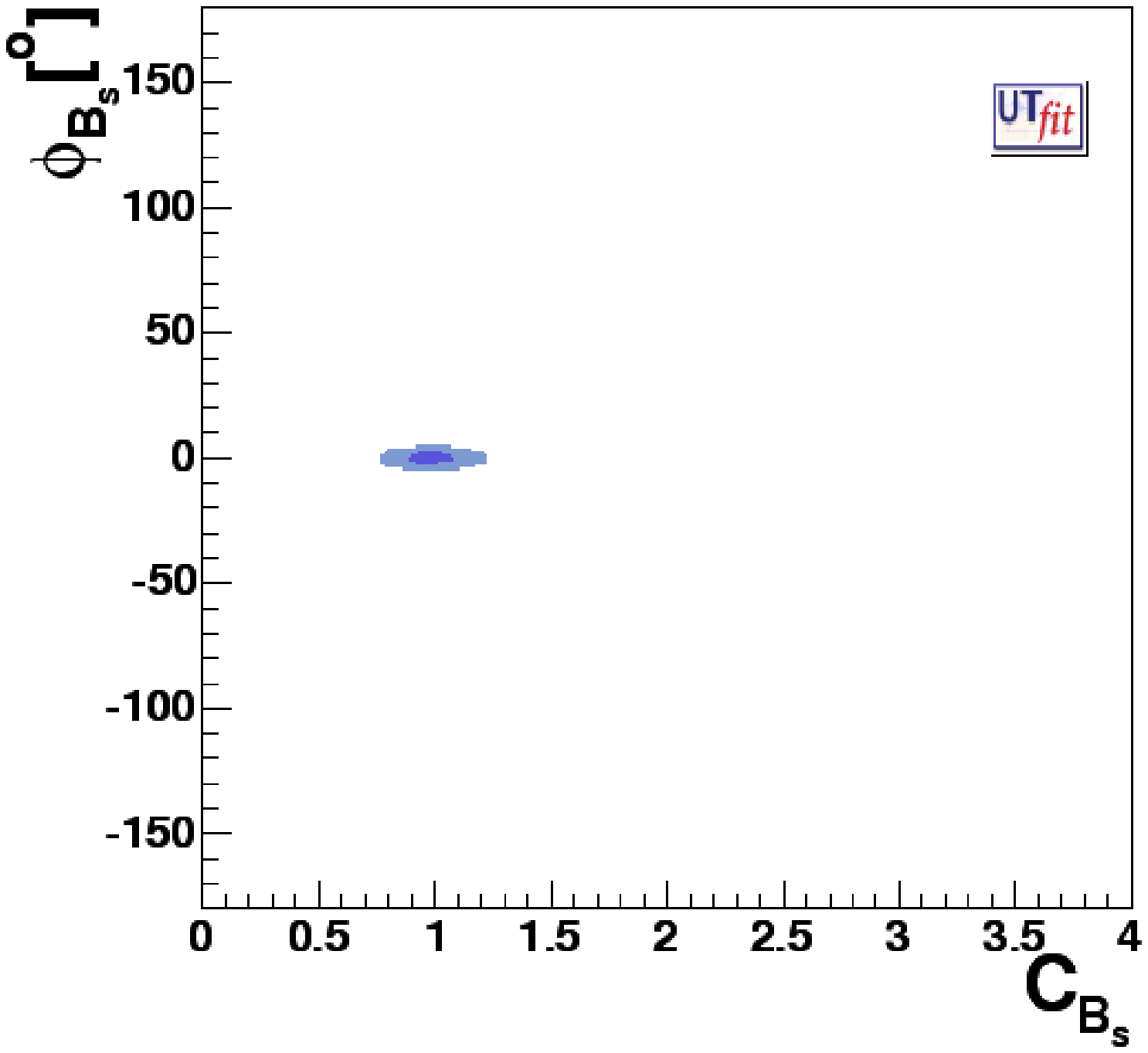} 
  \end{tabular}
  \caption{Outlook for Lepton-Photon 2009: the SM UT (top left), the
    UUT (top right), the $\phi_{B_d}$ vs. $C_{B_d}$ plane (bottom
    left) and the $\phi_{B_s}$ vs. $C_{B_s}$ plane (bottom right). See
    the text for details.}
  \label{fig:2009}
\end{figure*}

\section{Conclusions and Outlook}
\label{sec:concl}

Let us summarize the results presented in this talk in four messages:\\
\textbf{1} The recent results from $B$ factories make the UT fit
  overconstrained. This allows us to simultaneously fit SM CKM
  parameters and NP contributions to $\Delta F=2$ transitions, in the
  most general scenario with NP also affecting $\Delta F=1$ decays.
  With present data, the SM-like solution in the first quadrant for
  the UT is strongly favoured (see Fig.~\ref{fig:NP}). The nonstandard
  solution in the third quadrant has only $7$ \% probability.\\
\textbf{2} From the generalized UT analysis, we can conclude that NP
  contributions to $\Delta B = 2$ transitions can be of
  $\mathcal{O}(1)$ if they carry the same weak phase of the SM,
  otherwise they have to be much smaller or vanishing (see
  Fig.~\ref{fig:achilleplot}). New sources of flavour and CP violation
  must therefore be either absent (MFV) or confined to $b \to s$
  transitions. The latter possibility is naturally realized in many NP
  scenarios.\\
\textbf{3} In MFV models, the UUT can be determined, independently of NP
  contributions, with an accuracy comparable to the SM
  analysis. Together with the available data on $B \to X_s \gamma$, $B
  \to X_s l^+ l^-$ and $K^+ \to \pi^+ \nu \bar \nu$, this allows to
  derive stringent upper bounds on other rare $K$ and $B$
  decays. Sizable enhancements with respect to the SM are excluded,
  while strong suppressions are still possible at present.\\
\textbf{4} Although the constraints from $B \to X_s \gamma$ and $B \to X_s
  l^+ l^-$ are becoming more and more stringent, NP in $b \to s$
  transitions is still allowed to a large extent and might produce
  sizable deviations from the SM in the time-dependent CP asymmetries
  in $b \to s$ nonleptonic decays and in $B_s - \bar B_s$ mixing. This
  situation can be realized in SUSY models, where detailed
  computations of the deviations from the SM can be performed. 

We are bound to witness further improvements in the
experimental and theoretical inputs to the above analysis in the near
future. In the next few years, the UUT analysis might well become the
standard analysis, NP contributions to $\Delta F=2$ transitions will
be either revealed or strongly constrained, and rare decays will
provide stringent bounds on NP in $\Delta F=1$ processes or,
hopefully, show some deviation from the SM expectation. In
Fig.~\ref{fig:2009} I show a pessimistic view of what we might see at
Lepton-Photon 2009, in the dull scenario in which everything remains
consistent with the SM.\cite{utfitNP} Also in this case, however,
flavour physics will remain a crucial source of information on the
structure of NP. This information is complementary to the direct
signals of NP that we expect to see at the LHC.

I conclude reminding the reader that, for reasons of space, I had to
omit several very interesting topics, including in particular lepton
flavour violation and electric dipole moments, which might also reveal
the presence of NP in the near future.

\section*{Acknowledgments}
I am very much in debt to my flavour collaborators:
C. Bobeth, M. Bona, A.J. Buras, M. Ciuchini, T. Everth, E. Franco,
V. Lubicz, G. Martinelli, A. Masiero, F. Parodi, M. Pierini, P. Roudeau,
C. Schiavi, A. Stocchi, V. Vagnoni, S. Vempati, O. Vives and A. Weiler. I
acknowledge the support of the EU network "The quest for unification"
under the contract MRTN-CT-2004-503369.

\end{document}